\documentclass[aps,prc,twocolumn,superscriptaddress,nofootinbib]{revtex4-2}

\usepackage{url}
\usepackage{amsmath}
\usepackage{varwidth}
\usepackage{multirow}
\usepackage{dcolumn}
\usepackage{tabularx}
\usepackage{booktabs}
\newcolumntype{C}{>{\centering\arraybackslash}X}

\usepackage{graphicx}
\usepackage{makecell}
\usepackage{hyperref}
\hypersetup{
    colorlinks=true,
    linkcolor=blue,
    citecolor=blue,
    urlcolor=blue,
    anchorcolor=blue}

\begin{document}
\title{An introduction to the parton and hadron cascade model PACIAE 3.0}

\author{An-Ke Lei}
\affiliation{Key Laboratory of Quark and Lepton Physics (MOE) and Institute of
            Particle Physics, Central China Normal University, Wuhan 430079,
            China}

\author{Yu-Liang Yan}
\email[]{yanyl@ciae.ac.cn}
\affiliation{China Institute of Atomic Energy, P. O. Box 275 (10), Beijing
            102413, China}

\author{Dai-Mei Zhou}
\email[]{zhoudm@mail.ccnu.edu.cn}
\affiliation{Key Laboratory of Quark and Lepton Physics (MOE) and Institute of
            Particle Physics, Central China Normal University, Wuhan 430079,
            China}

\author{Zhi-Lei She}
\affiliation{School of Mathematical and Physical Sciences, Wuhan Textile
            University, Wuhan 430200, China}

\author{Liang Zheng}
\affiliation{School of Mathematics and Physics, China University of Geosciences
            (Wuhan), Wuhan 430074, China}

\author{Gao-Chan Yong}
\affiliation{School of Nuclear Science and Technology, University of Chinese
            Academy of Sciences, Beijing 100049, China}
\affiliation{Institute of Modern Physics, Chinese Academy of Sciences, Lanzhou
            730000, China}

\author{Xiao-Mei Li}
\affiliation{China Institute of Atomic Energy, P. O. Box 275 (10), Beijing
            102413, China}

\author{Gang Chen}
\affiliation{School of Mathematics and Physics, China University of Geosciences
            (Wuhan), Wuhan 430074, China}

\author{Xu Cai}
\affiliation{Key Laboratory of Quark and Lepton Physics (MOE) and Institute of
            Particle Physics, Central China Normal University, Wuhan 430079,
            China}

\author{Ben-Hao Sa}
\email[]{sabh@ciae.ac.cn}
\affiliation{Key Laboratory of Quark and Lepton Physics (MOE) and Institute of
            Particle Physics, Central China Normal University, Wuhan 430079,
            China}
\affiliation{China Institute of Atomic Energy, P. O. Box 275 (10), Beijing
            102413, China}

\date{\today}

\begin{abstract}
We introduce a parton and hadron cascade model PACIAE 3.0 based on
PYTHIA 6.428 and the PACIAE 2.2 program series. The simulation framework of
C-, B-, and A-loops are designed for the high energy
($\sqrt{s_{NN}}\geq 3$ GeV) and low energy ($\sqrt{s_{NN}}<3$ GeV) nuclear
collisions, respectively, in PACIAE 3.0. In the C-loop simulation, the
parton-parton inelastic scattering processes are added in the partonic rescattering
process. The single string structure and multiple string interaction mechanism
have been introduced investigating the strangeness enhancement
in C- and B-loop. An improved mapping relation between the centrality
percentage definition and the impact parameter definition is proposed
responding the observation of $b_{max}\approx 20$ fm from ALICE, ATLAS, and
CMS collaborations. We have extensively modified the phenomenological
coalescence hadronization model. The PACIAE 3.0 model simulated results of
particle yield, transverse momentum distribution, and rapidity distribution
well reproduce, respectively, the experimental data measured at FOPI, E895,
RHIC, and LHC energies.
\end{abstract}

\maketitle

\section{Introduction}
The phenomenological model-based Monte Carlo simulation is a powerful
tool to investigate the relativistic nuclear collisions and Quark Gluon Plasma
(QGP) phase transition observed there. To the end, various models have been
developed, such as PYTHIA~\cite{PYTHIA,pythia83}, HERWIG~\cite{herwig},
SHERPA~\cite{sherpa}, PCM~\cite{pcm}, HIJING~\cite{hijing}, QGSM~\cite{qgsm},
UrQMD~\cite{urqmd}, AMPT~\cite{ampt}, PACIAE~\cite{sa2},
THERMINATOR~\cite{therminator}, PHSD~\cite{phsd}, EPOS-LHC~\cite{eposlhc},
SMASH~\cite{smash}, JETSCAPE~\cite{jetscape} and Angantyr~\cite{angantyr} in
the high energy sector. At low energy the BUU-like models (such as BLOB,
BUU-VM, DJBUU, GiBUU, IBL, IBUU, LBUU, pBUU, PHSD, RBUU, RVUU, SMASH, SMF,
$\chi$BUU) and the QMD-like models (e.g., AMD, AMD+JAM, BQMD, CoMD, ImQMD,
IQMD-BNU, IQMD-SINAP, JAM, JQMD, LQMD, TuQMD/dcQMD, UrQMD) are
developed, cf.~\cite{zhang2018,ppnp2022,yong} and references therein.

PACIAE 3.0 is a parton and hadron cascade phenomenological model based on
PYTHIA~\cite{PYTHIA} and PACIAE 2.2
series~\cite{sa2,zhou,yan,She:2022vco,yan1}.
PACIAE model is developed from the LUCIAE~\cite{sa1,tai1,andersson} and
JPCIAE~\cite{jpciae} models. LUCIAE model was based on the FRITIOF~\cite{hpi}
with the extension of implementing both the Firecracker model (collective
multi-gluon emission in the interacting string color field) and the hadronic
rescattering. JPCIAE model was based on the JETSET and PYTHIA~\cite{torbi}
being able to simulate the relativistic hadron-hadron and heavy-ion collisions.
Soon after, the JETSET had been blended in PYTHIA, the JPCIAE was renamed as
PACIAE 1.0 correspondingly. As quoted in~\cite{torbi1}, not only
the LUCIAE model but also the JPCIAE (PACIAE 1.0) and even PACIAE 3.0 model
are all based on LUND String Fragmentation (LSF) regime.

In PACIAE 3.0 the C-, B- and A-simulation loops are designed
for the high energy ($\sqrt{s_{NN}}\geq3$ GeV) and low energy
($\sqrt{s_{NN}}< 3$ GeV) nuclear collisions, respectively. In all the
simulation loops the basic building block is a hadron-hadron ($hh$)
collision. The high energy $hh$ collision is a large momentum transfer and
small spatial scale process. It should be described first in the partonic
degree of freedom and then hadronized into the hadronic degree of freedom of
the final hadronic state by interfacing to PYTHIA~\cite{PYTHIA}. The low energy
$hh$ collision can be dealt with the elastic and inelastic two-body
scattering kinematics in hadronic degree of freedom only.

A couple of improvements in physics are introduced in PACIAE 3.0. They are
listed as follows:
\begin{itemize}
\item[(1)] The $hh$ total cross section is assumed to be proportional to the
nucleon-nucleon ($NN$) total cross section with coefficient equal to the ratio
of effective valence quark number in $hh$ collision system to that in $NN$
collision system~\cite{aqm,torbi1}. And the experimentally measured $NN$ total
cross section~\cite{alice1} is adopted.
\item[(2)] Three inelastic parton-parton scattering processes are added at
partonic rescattering stage in the C-loop simulation.
\item[(3)] Two strangeness enhancement mechanisms of single string structure
and multiple string interaction are implemented in the B- and C-loop
simulations.
\item[(4)] An improved mapping relation between percentage centrality
definition and impact parameter centrality definition responding the ALICE,
ATLAS, and CMS observation of $b_{max}\approx 20$ fm~\cite{sa2,yan1} is
proposed.
\item[(5)] The phenomenological coalescence hadronization model is extensively
modified in the C-loop simulation.
\end{itemize}

The PACIAE 3.0 program is now available on the open source platforms
GitHub and Gitee~\footnote{ \url{https://github.com/ArcsaberHep/PACIAE}; \\
\url{https://gitee.com/arcsaberhep/PACIAE}. }.

\section{Cumulative superposition of hadron-hadron collisions} \label{sec:hcs}
\begin{figure}[htbp]
\centering
\includegraphics[width=0.48\textwidth]{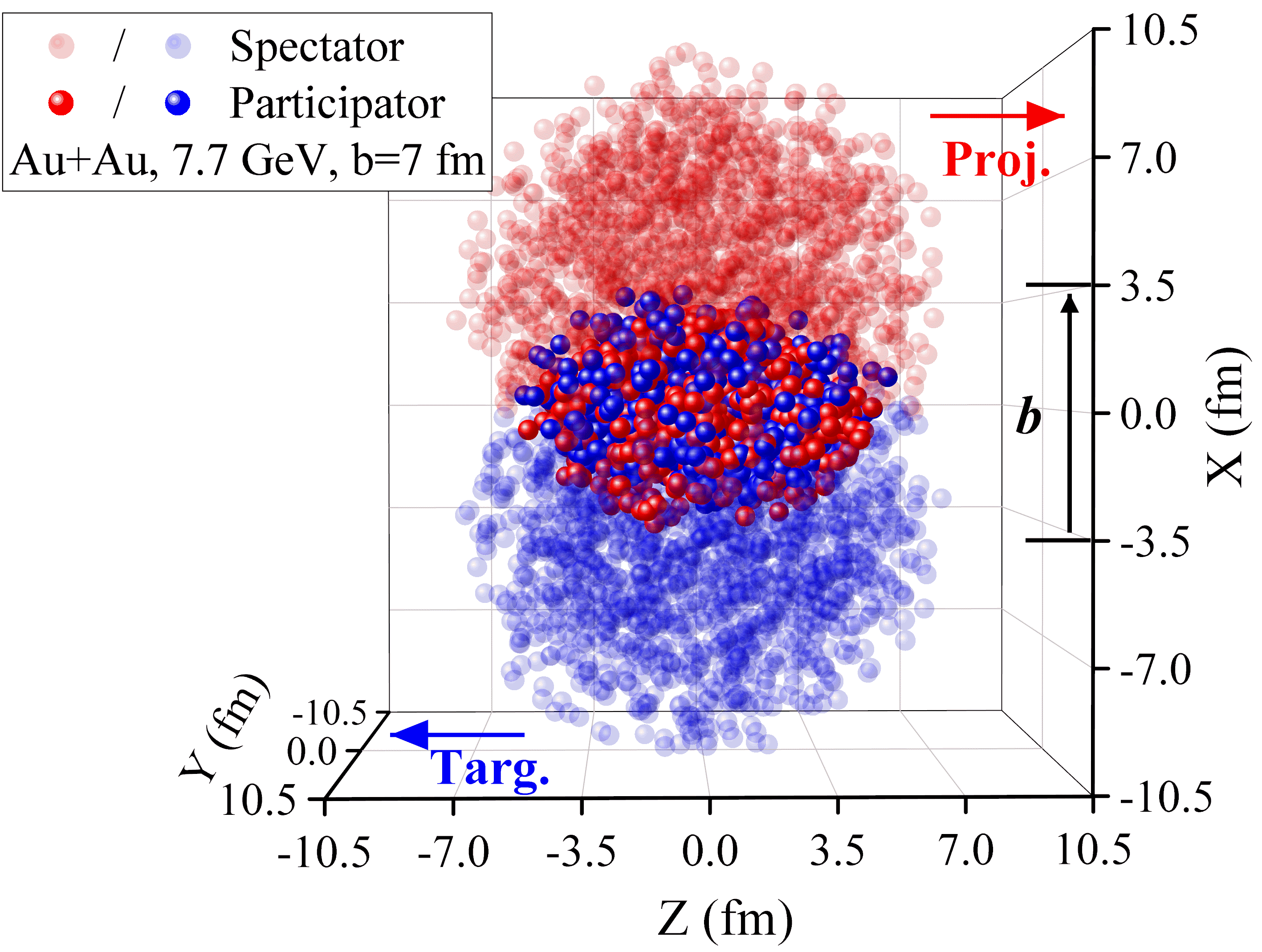}
\caption{The initial spatial distribution of nucleons in the impact
parameter $b=7$ fm Au+Au collisions at $\sqrt{s_{NN}}$ = 7.7 GeV.}
\label{r1}
\end{figure}

To begin with a heavy-ion collision simulation one first distributes the
nucleons in its own nucleus sphere by the Woods-Saxon distribution (for radius
$r$) and the uniform distribution in 4$\pi$ solid angle (for direction), as
shown in Fig.~\ref{r1}. Here the time origin is set at the moment of two
centers of the projectile and target spheres have the same coordinate of
$z=0$~\cite{sa2,sa,yan}.

\begin{figure}[htbp]
\centering
\includegraphics[width=0.48\textwidth]{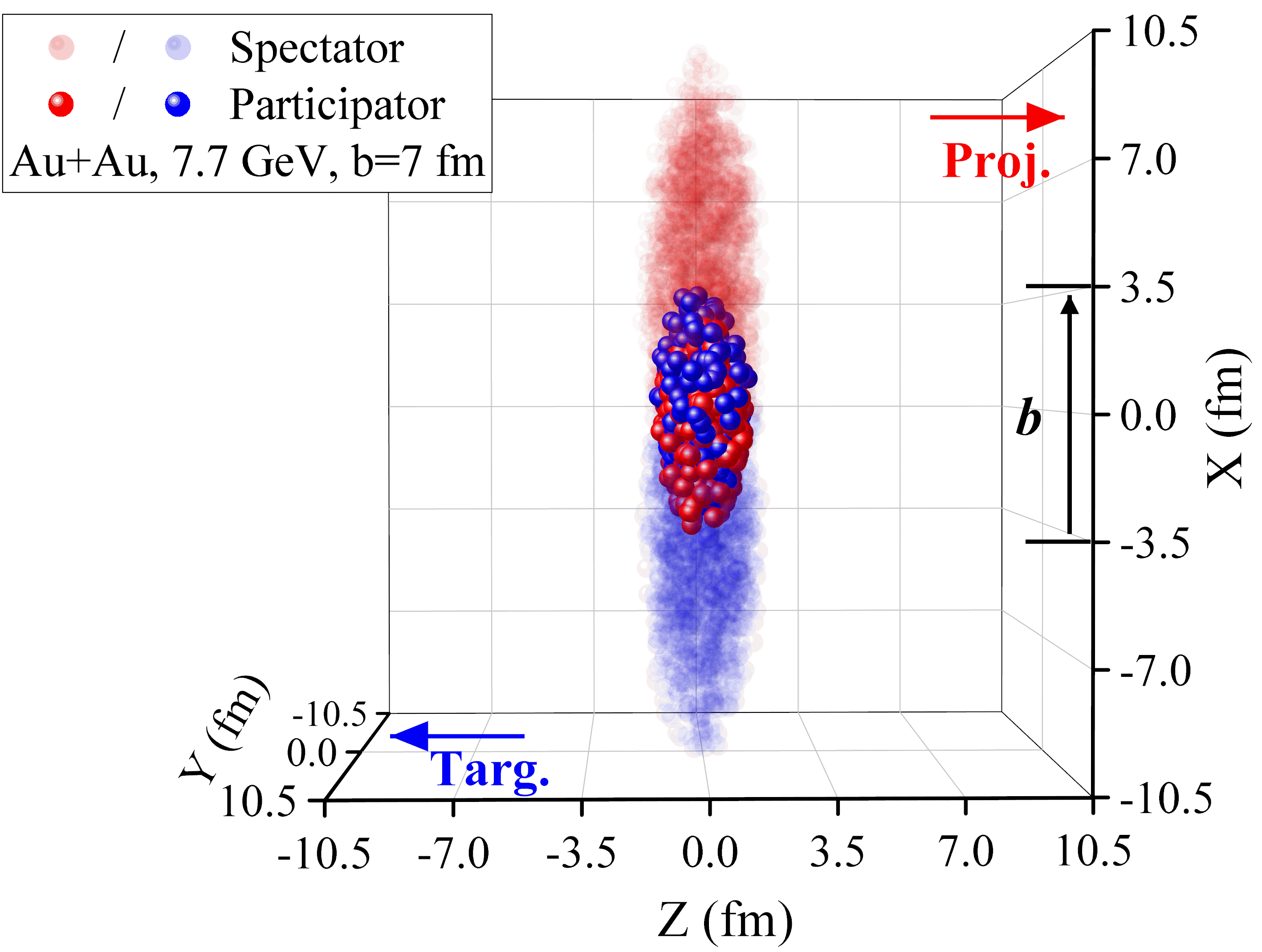}
\caption{The initial spatial distribution of nucleons in the impact
         parameter $b=7$ fm Au+Au collisions at $\sqrt{s_{NN}}$ = 7.7 GeV
         after Lorentz contraction.}
\label{r1afl}
\end{figure}

Takeing the Au+Au collision at $\sqrt{s_{NN}}$ = 7.7 GeV with the impact
parameter $b=7$ fm as an example, the initial momentum of each nucleon in the
projectile nucleus (Proj.) is $p_{x}=p_{y}= 0$ and $p_{z}=p_{\rm beam}$, and is
$p_{x}=p_{y}= 0$ and $p_{z}=-p_{\rm beam}$ in the target nucleus (Targ.). The
Lorentz contraction is then performed. Figs.~\ref{r1} and \ref{r1afl} show the
initial spatial distribution of nucleons before and after Lorentz contraction,
respectively. Meanwhile, the initial particle list, composed of four spatial
and four momentum vectors of all nucleons in the Au+Au collision system, is
constructed.

We assume the nucleon trajectory in the velocity field of nuclear collision
system is a straight line. A nucleon $i$ from projectile nucleus and $j$ from
target nucleus may collide if their relative transverse distance, $D$,
satisfies
\begin{equation}
D\leq\sqrt{\sigma_{NN}^{tot}/\pi},
\end{equation}
where $\sigma_{NN}^{tot}$ refers to the $NN$ total cross section. The
collision time $t_{ij}$ is then calculated~\cite{sa2}.

Two circulation loops are set: one for $i$ cycling over all the projectile
nucleons, another one for $j$ cycling over all the target nucleons. With the
calculated collision time $t_{ij}$ of all $i\text{-}j$ pairs the initial
$NN$ collision time list is constructed for a heavy-ion collision system.

A $NN$ collision with the least collision time is selected from the list. If it
is properly executed (see next section) its final hadronic state is
available and the generated hadrons are counted as its contribution to the
final hadronic state of the heavy-ion collision. The particle (nucleon or
hadron) list is then updated by removing two colliding particles from the
particle list and adding the generated particles to the particle list.
Consequently, the $NN$ ($hh$) collision time list is updated by removing the
$NN$ ($hh$) collision pair containing any one of the colliding particles from
the old collision time list and adding the new collision pairs composed of one
particle from the generated particles and another one from the old particle
list.

A new $NN$ ($hh$) collision with least collision time is then selected from the
updated collision time list and properly executed. With repeating the
aforementioned steps until the particle collision time list is empty, a Monte
Carlo simulation for a heavy-ion collision is finished.

\begin{figure}[htbp]
\centering
\includegraphics[width=0.28\textwidth]{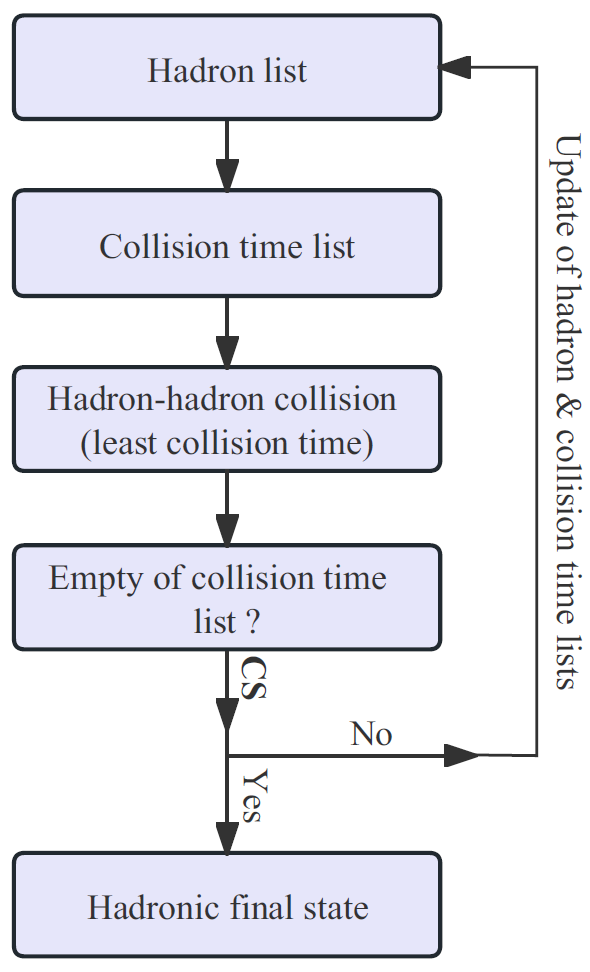}
\caption{A sketch for superposition of hadron-hadron collisions.}
\label{sket1}
\end{figure}

Therefore, in PACIAE model, a heavy-ion collision is indeed described as a
Cumulative Superposition (CS) of the $NN$ ($hh$) collisions, i.e. the
generated new hadrons will join in the processes of updating hadron list and
$hh$ collision time list, as shown in Fig.~\ref{sket1}.

\begin{figure}[htbp]
\centering
\includegraphics[width=0.45\textwidth]{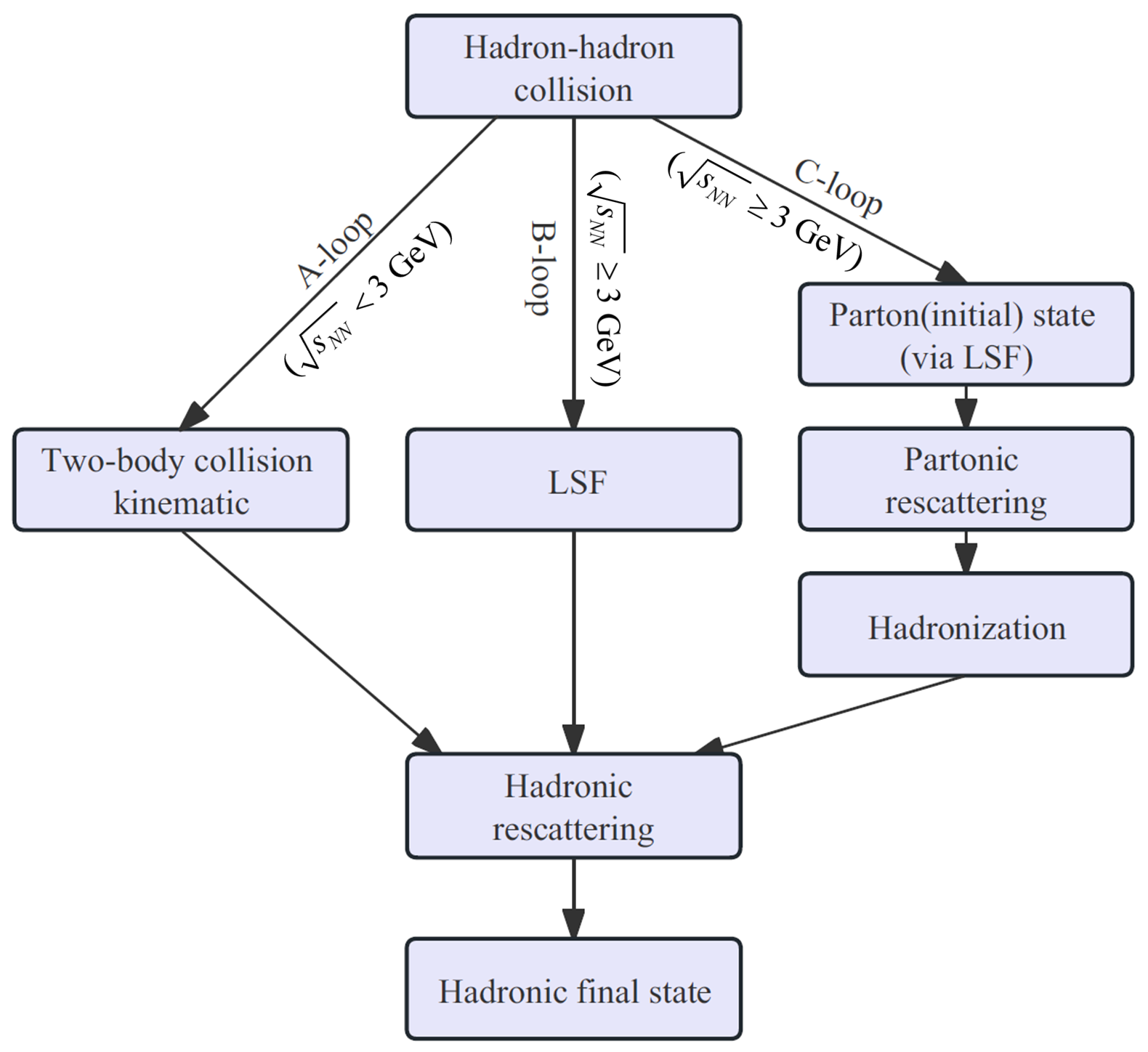}
\caption{A sketch of the hadron-hadron collision.}
\label{sket2}
\end{figure}

\section{Model for hadron-hadron collision} \label{sec:hh_collison}
The last section is common for the A-, B-, and C- simulation loops but leaves
a problem of the $hh$ collision execution. It will be addressed in this section
for A-, B-, and C-loops, individually.

The $hh$ collision in A-loop simulation is well described by the two-body
elastic and inelastic scattering kinematics in hadronic degree of
freedom~\cite{pythia83}, as shown in the left part of Fig.~\ref{sket2}.

Upto the second time of updating collision list, the inelastic scattering
is restricted to the following processes:
\begin{eqnarray*}
p + p \rightarrow \Delta^+ + p, \hspace{1.8cm}
p + p \rightarrow \Delta^{++} + n, \\
p + n \rightarrow \Delta^+ + n, \hspace{2cm}
p + n \rightarrow \Delta^0 + p, \\
n + n \rightarrow \Delta^0 + n, \hspace{2cm}
n + n \rightarrow \Delta^- + p, \\
\Delta^+ + p \rightarrow p + p, \hspace{2cm}
\Delta^+ + n \rightarrow p + n, \\
\Delta^0 + p \rightarrow p + n, \hspace{2cm}
\Delta^0 + n \rightarrow n + n, \\
\Delta^{++} + n \rightarrow p + p, \hspace{2cm}
\Delta^- + p \rightarrow n + n, \\
\pi^- + p \rightarrow \Delta^- + \pi^+, \hspace{2cm}
\pi^- + p \rightarrow \rho^0 + n, \\
\pi^- + p \rightarrow \rho^- + p, \hspace{1.5cm}
\pi^- + p \rightarrow \Delta^+ + \pi^-, \\
\pi^- + p \rightarrow \Delta^0 + \pi^0, \hspace{1.5cm}
\pi^- + n \rightarrow \Delta^- + \pi^0, \\
\pi^- + n \rightarrow \rho^- + n, \hspace{1.5cm}
\pi^- + n \rightarrow \Delta^0 + \pi^-, \\
\pi^+ + p \rightarrow \Delta^{++} + \pi^0, \hspace{1.5cm}
\pi^+ + p \rightarrow \Delta^+ + \pi^+, \\
\pi^+ + p \rightarrow \rho^+ + p, \hspace{1.3cm}
\pi^+ + n \rightarrow \Delta^{++} + \pi^-, \\
\pi^+ + n \rightarrow \Delta^0 + \pi^+, \hspace{1.5cm}
\pi^+ + n \rightarrow \Delta^+ + \pi^0, \\
\pi^+ + n \rightarrow \rho^0 + p, \hspace{2cm}
\pi^+ + n \rightarrow \rho^+ + n. 
\end{eqnarray*}

For both the elastic and inelastic scattering processes, the four momenta of
scattered hadrons are determined by the energy-momentum conservation
\cite{sa2}. Among the inelastic scattering processes, if it is an exothermic
reaction, such as $p + p \rightarrow \Delta^+ + p$, the threshold energy
effect is taken into account. For an exothermic inelastic scattering, if the
kinetic energy of its incident channel is less than the threshold energy, it
should be dealt as an elastic scattering rather than inelastic scattering
originally. Here two parameters are essential: One is the ratio of
inelastic cross section to total cross section $R_{inela/tot}$ (`x\_ratio' in
program). Another is the $\Delta$ particle instantaneously decay probability
(`decpro') at the moment of formation.

Inspired by the additive quark model \cite{aqm}, we assume different outgoing
channels and the resonance production process developed from a given incident
channel are equally distributed. Upto the second time of updating collision
list, there is only one resonance process of $p+\pi^+\rightarrow\Delta^{++}$
to be considered.

In PACIAE model, the experimental data of $\sigma_{NN}^{tot}\approx 70$ mb
measured at LHC energies and $\sigma_{NN}^{tot}\approx 40$ mb measured at
RHIC energy and below are adopted~\cite{alice1}. The total cross section of
$IJ$ collision (hadron $I$ bombards with $J$), is assumed to be proportional
to the $NN$ collision one, with the coefficient calculated
by~\cite{aqm,pythia83}:
\begin{equation}
C_{IJ}=\frac{n_{eff}^In_{eff}^J}{n_{eff}^Nn_{eff}^N},
\label{coef}
\end{equation}
\begin{equation}
n_{eff}^I=n_d^I+n_u^I+0.6n_s^I+0.2n_c^I+0.07n_b^I.
\end{equation}
In above equation the $n_i^I$, refers to the number of
effective $i$-th valence quark (antiquark) in the $I$-th hadron.

\begin{figure*}[htbp]
\centering
\includegraphics[width=0.90\textwidth]{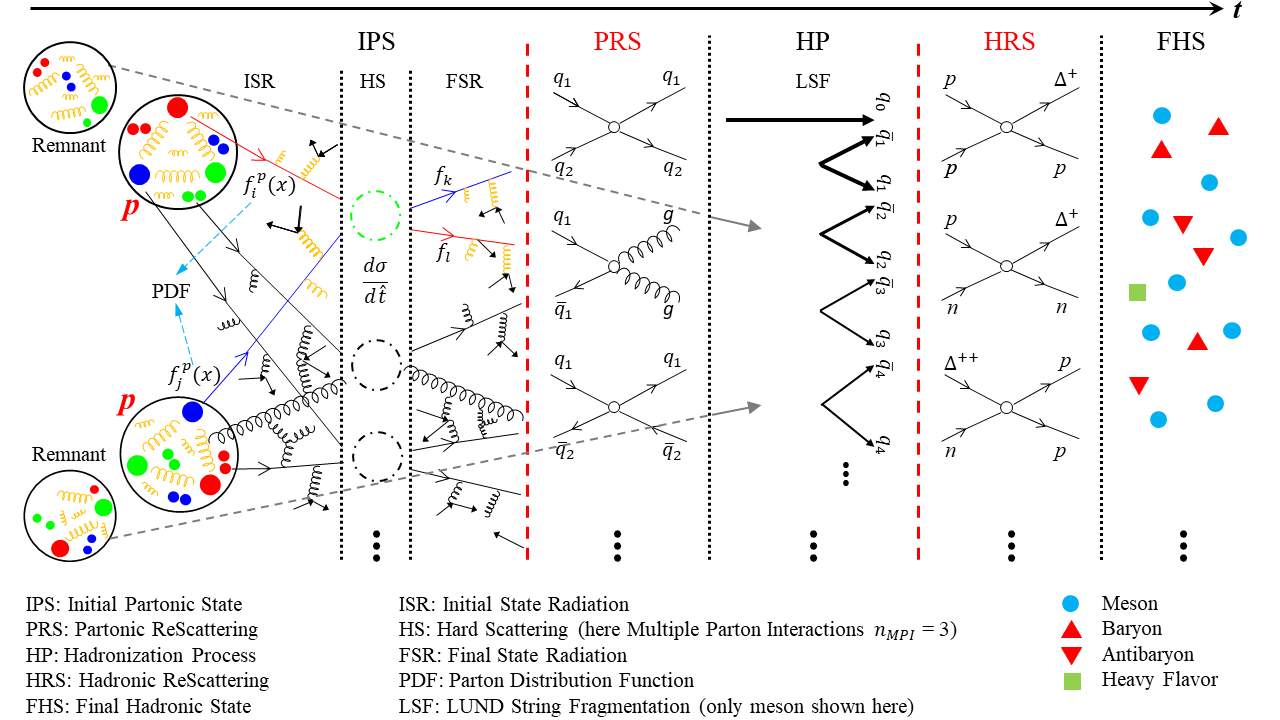}
\caption{A sketch for the physical routines in a high energy $pp$ simulation.}
\label{phyrou}
\end{figure*}

Differently, in the high energy B-loop simulation the final hadronic state of
a $hh$ collision is supplied by PYTHIA~\cite{PYTHIA}: As a proton consists of
three valence quarks, countless sea quarks and gluons, a $pp$ collision,
may comprise $n_{MPI}$ parton-parton pair
interactions. Here $n_{MPI}$ refers to the number of MultiParton Interactions
(MPI). Each parton-parton collision is described by a Hard Scattering
(HS) together with the Initial State Radiation (ISR, or initial state parton
shower) and Final State Radiation (FSR, or final state parton shower). The
resulted partons then hadronize together with two remnants providing the final
hadronic state for a $pp$ collision. Here the remnant refers to the left part
of colliding proton, beside the ones join in the hard scattering.
Fig~\ref{phyrou} without PRS (Partonic ReScattering) and HRS
(Hadronic ReScattering) is just a schematic diagram of the physical processes
included in a $pp$ collision.

\begin{figure}[htbp]
\centering
\includegraphics[width=0.45\textwidth]{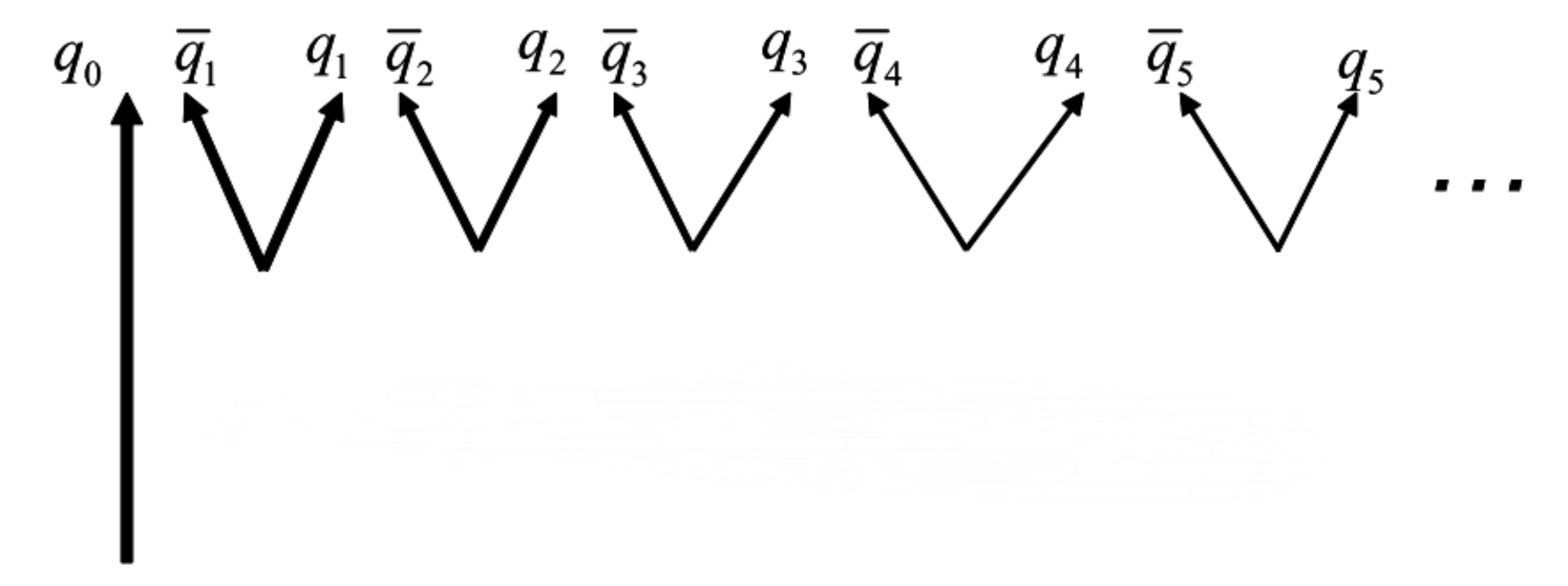}
\caption{The Feynman diagram like sketch for the string iterative breaking
    processes starting from the quark end of a $q_0\bar q_0$ string.}
\label{qbra}
\end{figure}

The hadronization in PYTHIA model is phenomenologically described by the
string iterative breaking processes: In case of the iterative string breaking
process starts at the $q_0$ end of a $q_0\bar q_0$ string, if the string
energy is large enough, a new $q_1\bar q_1$ pair may be excited from the
vacuum, such that a meson of $q_0\bar q_1$ may formed and left behind the
quark $q_1$. Later on, $q_1$ quark in its turn may excite a $q_2\bar q_2$ pair
from the vacuum and combines another meson together with the $\bar q_2$.
Repeating this breaking process, a lot of mesons are formed in the final
hadronic state of the $hh$ collision system, as shown in Fig.~\ref{qbra}.

\begin{figure}[htbp]
\centering
\includegraphics[width=0.45\textwidth]{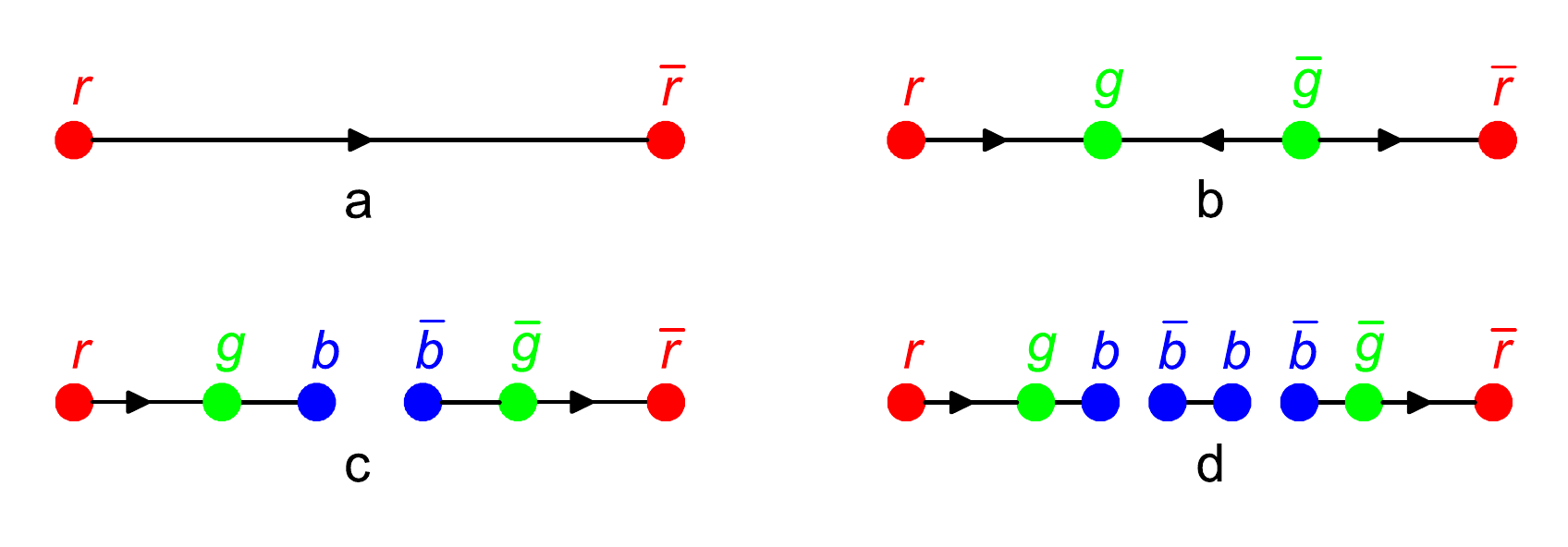}
\caption{The step-wise sketch illustrating the popcorn production of a
         baryon-antibaryon pair in the string iterative breaking processes,
         taken from \cite{pythia83}.}
\label{barpro}
\end{figure}

Fig.~\ref{barpro}, taken from Ref.~\cite{pythia83}, shows the baryon (antibaryon)
generation process in the popcorn model~\cite{pythia83}: One starts from a
red-antired ($r\bar r$) string (with color flow indicated by the arrow in
panel a). A green-antigreen ($g\bar g$) pair may be excited from vacuum
between $r\bar r$ reversing the color flow in the central part of the string
(panel b). A third blue-antiblue ($b\bar b$) pair is created and breaks the
string into two (panel c). Then another string breaking process happens and
produces a $b\bar b$ meson between the baryon ($rgb$) and antibaryon
($\bar b\bar g\bar r$).

Takeing meson production as an example, once the $q_{i\text{-}1}$ and
$\bar q_i$ flavors are sampled, a selection should be made between the
possible multiplets. The different multiplets have different relative
composition probability, which is not given by first principle but must
depend on the fragmentation processes, cf. \cite{PYTHIA} for the details.

In C-loop $hh$ collision simulation, we first forcedly turn-off the
hadronization before the execution of PYTHIA and break-up the strings and
diquarks after the execution of PYTHIA, resulting an initial partonic state.
This partonic state then undergoes the partonic rescattering, where the
lowest-order perturbative quantum chromodynamics (LO-pQCD) parton-parton
interaction cross section~\cite{Combridge,Field} is employed. After partonic
rescattering the hadronization is implemented by the LUND string
fragmentation regime and/or the coalescence model (see Sec.~\ref{sec:coal}).
The hadronic rescattering is then followed, generating a final hadronic state
for a $hh$ collision system. Meanwhile, this simulation could be selected to
stop at any stage desired conveniently. Fig.~\ref{phyrou} shows
the above physical processes in a C-loop $pp$ simulation.

\section{Partonic rescattering in C-loop simulation} \label{sec:prs}
In PACIAE model, the partonic rescattering is implemented in C-loop simulation
and only $2\rightarrow 2$ processes are considered. The simulation framework of
partonic rescattering is similar to that in the Sec.~\ref{sec:hcs}: We
first construct an initial parton-parton collision time list based on the
parton list in the initial partonic state. Secondly, a parton-parton collision
with least collision time is performed. Thirdly, the parton list and
parton-parton collision time list are updated. A new parton-parton collision
with least collision time is then selected from the updated collision time
list and properly executed. With repeating the aforementioned steps until the
parton-parton collision time list is empty, a Monte Carlo simulation for
partonic rescattering is finished.

Table~\ref{tab1} gives the considered parton-parton interactions, where the
differential cross section is expressed in the form of
\begin{equation}
\frac{d\sigma}{dt}(ab\rightarrow cd;s,t)=K\frac{\pi\alpha_s^2}{s^2}
|\overline M(ab\rightarrow cd)|^2,
\label{qqcro}
\end{equation}
and calculated by the LO-pQCD approximation~\cite{Combridge,Field}. In the
equation above the $\alpha_s$ refers to strong coupling factor. The $s$, $t$
and $u$ (cf. Table~\ref{tab1}) are the Mandelstam invariants in the
kinematics of $ab\rightarrow cd$ quark process. And $K$ is an enlarged
factor introduced empirically. The corresponding integral cross section is
\begin{equation}
\frac{d\sigma}{dt}(ab\rightarrow cd;s)=
\int_{-s}^0\frac{d\sigma}{dt}(ab\rightarrow cd;s,t)dt.
\end{equation}
As the differential cross section is divergent at $t\rightarrow 0$, Debye
screening coefficient $\mu$ has to be introduced. Therefore, taking the number
1 process in Table~\ref{tab1} as an example, its matrix element in
differential cross section should be modified to
\begin{equation}
|M(q_1q_2\rightarrow q_1q_2)|^2=\frac{4}{9}\frac{s^2+u^2}{t^2-\mu^2}.
\end{equation}

\begin{table*}[tbp]
\caption{Parton-parton collisions.}
\setlength{\tabcolsep}{0.11\textwidth}
\renewcommand{\arraystretch}{1.8}
\begin{tabular}{c|c|c}
\hline
\hline
Order& Process& $|\overline M|^2$ \\
\hline
 1& $q_1q_2\rightarrow q_1q_2$& $\frac{4}{9}\frac{s^2+u^2}{t^2}$ \\
\hline
 2& $q_1q_1\rightarrow q_1q_1$& $\frac{4}{9}(\frac{s^2+u^2}{t^2}+
        \frac{s^2+t^2}{u^2})-\frac{8}{27}\frac{s^2}{ut}$ \\
\hline
 3& $q_1\bar q_2\rightarrow q_1\bar q_2$& $\frac{4}{9}\frac{s^2+u^2}{t^2}$ \\
\hline
 4& $q_1\bar q_1\rightarrow q_2\bar q_2$& $\frac{4}{9}\frac{t^2+u^2}{s^2}$ \\
\hline
 5& $q_1\bar q_1\rightarrow q_1\bar q_1$& $\frac{4}{9}(\frac{s^2+u^2}{t^2}+
        \frac{t^2+u^2}{s^2})-\frac{8}{27}\frac{u^2}{ts}$ \\
\hline
 6& $q\bar q\rightarrow gg$& $\frac{32}{27}\frac{u^2+t^2}{ut}-\frac{8}{3}
        \frac{u^2+t^2}{s^2}$ \\
\hline
 7& $gg\rightarrow q\bar q$& $\frac{1}{6}\frac{u^2+t^2}{ut}-\frac{3}{8}
        \frac{u^2+t^2}{s^2}$ \\
\hline
 8& $qg\rightarrow qg$& $-\frac{4}{9}\frac{u^2+s^2}{us}+\frac{u^2+s^2}{t^2}$ \\
\hline

 9& $gg\rightarrow gg$& $\frac{9}{2}(3-\frac{ut}{s^2}-\frac{us}{t^2}-
        \frac{st}{u^2})$ \\
\hline
\hline
\end{tabular}\label{tab1}
\end{table*}

Among the listed parton-parton collisions in Table~\ref{tab1}, the number
1, 2, 3, 5, 8, and 9 processes are elastic scattering processes. In the elastic
scattering process, as quark flavors in incident and outgoing channels
are unchanged, it is easy to handle. Most of the parton and hadron transport
models, like AMPT~\cite{ampt} and the PACIAE 2.2, only take elastic
parton-parton scattering processes into account. In PACIAE 3.0, the number 4, 6, and
7 inelastic parton-parton scattering processes are implemented.

In the number 4 and 7 inelastic scattering processes, if the invariant mass of
incident channel is large enough, the available outgoing flavor may be
different. We assume the different outgoing flavor is distributed
inversely proportional to the $x$-th power of its respective constituent quark
masse ($p_q\propto m_q^{-x}$). Here $x$ is a parameter (default, D=3.65).

\section{Hadronic rescattering}
The simulation framework of hadronic rescattering is also similar to the one
in the Sec.~\ref{sec:hcs}. However, here we first filter out
the desired hadrons from the available hadron list after hadronization to
construct an initial hadron list. We then construct a hadronic collision
time list, select a $hh$ collision pair with least collision time and execute
it properly, update hadron list and hadronic collision time list, etc., one
step after another, like that in the Sec.~\ref{sec:hcs}.

Here the $NN$ total cross section is also taken from experiment and the
total cross section of $IJ$ incident channel is assumed to be proportional
to the $NN$ one with coefficient given by Eq.~(\ref{coef}). The ratio of
inelastic to total cross section (`x\_ratio' in program, D=0.85) is a model
parameter, too.

In the hadronic rescattering we consider nearly 600 different inelastic
$hh$ collisions (cf. program packet hadcas\_30.f), besides the elastic $hh$
collision. The inelastic $hh$ collisions listed at the begin of section
\ref{sec:hh_collison}, for instance, are main parts of them. If user desired
channel is not in the 600 list, it has to be added manually.

\section{Centrality definition and expression}
In experiment, the centrality of a nucleus-nucleus ($AB$) collision is usually
defined as a percentile $c$ in the $AB$ total cross section and is
assumed to be approximately equivalent to the fraction of charged particle
multiplicity above a multiplicity cut of $N_{ch}^{cut}$~\cite{yan1}:
\begin{equation}
c\approx \frac{1}{N_{ch}^{tot}}\int_{N_{ch}^{cut}}^{N_{ch}^{tot}}
d\sigma/dN_{ch}^{'} dN_{ch}^{'}.
\label{bas3}
\end{equation}
Meanwhile, this percentile $c$ is also assumed to be equivalent to the
fraction in impact parameter distribution~\cite{sa2,Broniowski}
\begin{equation}
c\approx \frac{2}{b_{max}^2}\int_0^b b^{'}db^{'}.
\end{equation}
Therefore a mapping relation
\begin{equation}
b=\sqrt c \times b_{max},
\label{eq1}
\end{equation}
is obtained~\cite{sa2,yan1}. In the above equation $b_{max}$ is defined as
\begin{equation}
b_{max}=R_A+R_B+2\times d,
\end{equation}
where $R_A$, for instance, is given by
\begin{equation}
R_A=r_0A^{1/3}, \hspace{0.5cm} r_0=1.12 fm.
\end{equation}
Here $A$ also denotes the atomic number of the nucleus. $d=0.546$~fm
refers to the diffusivity parameter describing the tail of nuclear density
profile.

In response to the observations from ALICE, ATLAS, and CMS that the maximum
impact parameter should be extended to 20 fm, in PACIAE 3.0 we assume
\begin{equation}
b_{max}=R_{A}+ R_{B}+f\times d.
\label{eq2}
\end{equation}
The coefficient $f$ is fixed by fitting the results of improved Monte Carlo
Glauber model simulations~\cite{loiz}. It gives $f\approx 4$ for the
nucleus-nucleus collisions and $f\approx 2$ for the proton-nucleus
collisions~\cite{yan1}. Therefore we propose that, in the
absence of experimental data on impact parameters, one can employ the
Eqs. (\ref{eq1}), (\ref{eq2}) and fitted $f$ value to calculate it.

In the heavy-ion collisions, the centrality is always represented by the number
of participant nucleons $N_{part}$.  It is calculated, in PACIAE model, by the
optical Glauber model~\cite{sa2,yan1,esk}. A relationship of
\begin{equation}
\langle T_{AA}\rangle=\frac{\langle N_{coll}\rangle}{\sigma_{NN}^{inel}},
\label{tanc}
\end{equation}
is employed to calculate the binary $NN$ collision number $N_{coll}$. In the
above equation $T_{AA}$ is the nuclear overlap function and the angle bracket
indicates averaging over events.

\section{Strangeness enhancement}
In the string fragmentation picture of the relativistic $NN$ and heavy-ion
collisions, strange quark production is
suppressed comparing to up and down quarks due to the tunneling
probability~\cite{PYTHIA}
\begin{eqnarray}
P(m_{\perp q})=\exp(-\frac{\pi}{\kappa}m_q^2) \exp(-\frac{\pi}{\kappa}p_{\perp q}^2), \label{eqn:tunnel}
\end{eqnarray}
where the $\kappa\approx$ 1 GeV/fm $\approx$ 0.2 GeV$^2$ is the (vacuum)
string tension for a pure $q\bar q$ string. However a pronouncing enhancement
of strange particle relative to pion production is really observed by ALICE
collaboration in the relativistic $pp$ collisions~\cite{ALICEnp}. To this end,
we introduce an effective string tension stemming from single string
structure~\cite{Tai} and the multiple string interaction~\cite{zheng,zhoustr}
instead of (vacuum) string tension in Eq.~(\ref{eqn:tunnel}).

In Ref.~\cite{Tai}, we have constructed a parameterized effective
string tension coming from the single string structure:
\begin{equation}
\kappa_{eff}^{s}=\kappa_0 (1-\xi)^{-\alpha}.
\label{eqn:single_kappa}
\end{equation}
In the above equation, $\kappa_0$ is string tension of pure (dipole)
$q\bar q$ string. $\alpha$ is a parameter to be tuned with experimental
data. $\xi$ is parameterized as:
\begin{equation}
\xi = \frac{\ln(\frac{k_{\perp max}^2}{s_0})}{\ln(\frac{s}{s_0})+\sum_{j=gluon}\ln(\frac{k_{\perp j}^2}{s_0})},
\end{equation}
where $k_{\perp}$ denotes the transverse momentum of the gluons inside a dipole
string. The $\sqrt{s}$ and $\sqrt{s_0}$ give the mass of the string system and
the parameter related to the typical hadron mass, respectively.
The $\xi$ quantifies the difference between a gluon wrinkled string and a pure
$q\bar q$ string. The value of this effective string tension changes on a
string-by-string basis in the current implementation and takes the string-wise
fluctuations into consideration.

Later on, we consider the multiple string interaction effects from the
correlation of strings overlapping in a limited transverse space by
parameterizing the effective string tension, in a manner similar to the
close-packing strings discussed in Ref.~\cite{Fischer:2016zzs} as follows:
\begin{equation}
\kappa_{eff}^{m}=\kappa_0 (1+\frac{\frac{N_{coll}}{N_{part}}n_{MPI}-1}{1+p^{2}_{T\ ref}/p^{2}_0})^{r}.
\label{eqn:global_kappa-p}
\end{equation}
In the above equation, the $n_{MPI}$ indicates the number of multiple parton
interactions in a $pp$ collision system and $p^{2}_{T\ ref}/p^{2}_0$ shows the
transverse scale of a typical string object relative to the proton size. The
exponent $r$ is then treated as a free parameter. As larger
$n_{MPI}$ leads to a denser string system in an event, $n_{MPI}$
strongly correlates with the charged particle multiplicity. The factor of
$N_{coll}/N_{part}$ amplifies the multiple string interaction effects in
heavy-ion collisions~\cite{zhoustr}. Multiplying $\kappa^{s}_{eff}$ on both
side of Eq.~(\ref{eqn:global_kappa-p}), one obtains
\begin{eqnarray}
\kappa^{s}_{eff}\times\kappa^{m}_{eff}= & \kappa^{s}_{eff} (1+\frac{\frac{N_{coll}}{N_{part}}n_{MPI}-1}{1+p^{2}_{T\ ref}/p^{2}_0})^{r} \nonumber \\
     \equiv & \kappa^{s+m}_{eff}.
 \label{eqn:couple_kappa-p}
\end{eqnarray}

In PYTHIA~\cite{PYTHIA} the strange quark suppression relevant parameters are:
\begin{itemize}
    \item[(1)] PARJ(1), the suppression of diquark-antidiquark pair production
                        in string-breaking process, compared with
                        quark-antiquark pair production.
    \item[(2)] PARJ(2), the suppression of $s$ quark pair production compared
                        with $u$ or $d$ pair production.
    \item[(3)] PARJ(3), the extra suppression of $s$ diquark production
                        compared with the normal suppression of $s$ quarks.
    \item[(4)] PARJ(21), Gaussian width of the transverse momentum
                        distribution for primary hadrons in fragmentation.
\end{itemize}
They can be related to the effective string tension through a scaling function
implied by the tunneling probability:
 \begin{equation}
 \lambda_2=\lambda_1^{\frac{\kappa^{eff}_1}{\kappa^{eff}_2}}.
 \label{lamd}
 \end{equation}
In the above equation, $\kappa^{eff}_1$=1 GeV/fm represents the vacuum string
tension and $\kappa^{eff}_2$ is the effective string tension. The $\lambda_1$
and $\lambda_2$ refer to the one among PARJ(1), PARJ(2), and PARJ(3) before
and after modification, respectively. The $\lambda_2$ will be enlarged when
the effective string tension $\kappa^{eff}_2$ becomes greater than
$\kappa^{eff}_1$.

Similarly, the PARJ(21) varies with the effective string tension as
 \begin{equation}
 \sigma_2=\sigma_1(\frac{\kappa^{eff}_2}{\kappa^{eff}_1})^{1/2}.
 \end{equation}

\section{Phenomenological coalescence hadronization model} \label{sec:coal}
There are two hadronization mechanisms implemented in C-loop simulation:
The LUND string fragmentation regime and the coalescence (hadronization) model
COCCNU (CO: the moral of coalescence, CCNU: the short of Central China Normal
University). It is a phenomenological coalescence model unlike the
semi-analytical coalescence models in 
Refs.~\cite{yang,ko,Greco:2003mm,steff,Fries:2003kq,sandon}.

In the PACIAE C-loop simulation, if coalescence model is selected, one then
starts from the parton list (composed of quarks, antiquarks, and gluons)
available after partonic rescattering. All the gluons in this parton list are
randomly split into quark-antiquark pairs, resulting in a new parton list
composed of quarks (antiquarks) only.

Then the collision system proceeds with energetic quark (antiquark)
deexcitation process: A cycle over quark (antiquark) in the parton list is
constructed. If the energy of a quark (antiquark) is larger than the
deexcitation threshold energy $e_{she}$, it deexcites according to the
vacuum excitation regime of $q_0\rightarrow q_0q_1\bar q_1$
($\bar q_0\rightarrow \bar q_0q_1\bar q_1$)~\cite{PYTHIA}.
The generated quark-antiquark pair is filled at the end of the parton list.
This deexcitation process is continuously repeated until the quark (antiquark)
energy goes down to $e_{she}$. In each step, the transverse momenta of
generated quark-antiquark pair are sampled according to the Gaussian or
exponential distributions (controlled by parameter `i\_pT'). The generated
quark-antiquark pair takes a part of its mother quark (antiquark) energy, the
fraction of this part is sampled randomly from an uniform distribution or
fragmentation functions~\cite{PYTHIA} (controlled by parameter `adj1(29)').
Of course, the corresponding four momentum should be subtracted from mother
quark (antiquark). The finishing of this cycle presents the end of the first
generation deexcitation, followed by the second generation deexcitation with a
new cycle over the generated quarks (antiquarks). A free parameter
`adj1(16)' is set for the allowed maximum number of deexcitation generation
(D=1).

In the gluon splitting and the energetic quark (antiquark) deexcitation
processes, a key problem is the flavor generation probability of the outgoing
channel. We assume the different outgoing flavors are distributed
inversely proportional to the $x$-th power of their respective constituent
quark masses. Here the $x$ is a parameter. The $x$=1 is assumed
in~\cite{torbi1} for the calculation of effective number of quarks in the
hadron.

After the gluon splitting and the energetic quark (antiquark) deexcitation,
the collision system is represented by a quark (antiquark) list. Then it
proceeds to a combination loop: Selecting a proper quark and antiquark from
the parton list to form a specific meson in the meson Table \ref{Meson},
and/or choosing three quarks (antiquarks) to coalesce into a specific baryon
(antibaryon) in the baryon Table \ref{Baryon}. Here many strategies are
possible, for example, the combination starts from quark or antiquark, to
combine into a meson or baryon, etc. Which one is better has to be decided by
reproducing the experimental data. Presently, the combination starts from
antiquark in PACIAE 3.0. An antiquark is assumed to form an antibaryon together
with two other antiquarks by probability
$p=\frac{adj1(31)*adj1(33)}{1+adj1(31)*adj1(33)}$ and to form a meson together
with a quark by probability $(1-p)$. adj1(31) and adj1(33) are two free
parameters in the program. This combination loop is performed over the parton
list until it is empty. If the empty of parton list is hard to reach, the
remaining partons will attempt to re-hadronize by string
fragmentation~\cite{PYTHIA}.

We assumes the three-momentum of the coalesced hadron is the sum of its
constituent quark (antiquark) three-momentum. The extra energy (the part
deviated from the conservation) is additionally counted into a specific array,
left for sharing among partons and hadrons in the current list. The
three-position of the coalesced hadron is the random summation of the
three-position of its constituent quark (antiquark). The time of
coalesced hadron is assumed to be the latest time among the
constituent quarks (antiquarks).

Meanwhile, the phase space constraint
\begin{equation}
   \frac{16\pi^2}{9}\Delta r^3\Delta p^3=\frac{h^3}{d},
\end{equation}
is considered. In the above equation the $h^3/d$ is the volume occupied by a
single hadron in the phase space, $d$=4 refers to the spin and parity
degeneracies of the hadron. The $\Delta r$ and $\Delta p$ stand for the
sum of pair-wise relative distances between two (meson) or among three
(baryon) partons in the spatial and momentum phase spaces, respectively.

\begin{table*}[tbp]
\centering
\renewcommand{\arraystretch}{1.1}
\caption{Mesons in coalescence hadronization model.}
\begin{tabularx}{\textwidth}{C|C|C|C|C|C|C}
\hline
\hline
  & \multicolumn{3}{c|}{Pseudoscalar Meson} & \multicolumn{3}{c} {Vector Meson} \\
 \hline
 Quark conf. & Name & Mass(GeV) & Proper probability & Name & Mass(GeV) & Proper probability \\
\hline
$u\bar d$  & $\pi^+$   & 0.1396  & 1  & $\rho^+$  & 0.7669  &   1    \\
\hline
$d\bar u$  & $\pi^-$   & 0.1396  & 1  & $\rho^-$  & 0.7669  &   1    \\
\hline
$u\bar s$  & $K^+$   & 0.4936  & 1  & $K^{*+}$  & 0.8921  &   1    \\
\hline
$s\bar u$  & $K^-$   & 0.4936  & 1  & $K^{*-}$   & 0.8921  &   1    \\
\hline
$d\bar s$  & $K^0$   & 0.4977  & 1  & $K^{*0}$  & 0.8962  &   1    \\
\hline
$s\bar d$  & $\bar{K^0}$   & 0.4977  & 1  & $\bar{K}^{*0}$  & 0.8962  &   1    \\
\hline
$u\bar u$  & $\pi^0$   & 0.1350  & 0.5  & $\rho^0$  & 0.7700  &   0.5    \\
\hline
$u\bar u$  & $\eta$   & 0.5488  & 0.167  & $\omega$  & 0.7820  &  0.5   \\
\hline
$u\bar u$  & $\eta'$   & 0.9575  & 0.333  &    &    &       \\
\hline
$d\bar d$  & $\pi^0$   & 0.1350  & 0.5  & $\rho^0$  & 0.7700  &   0.5    \\
\hline
$d\bar d$  & $\eta$   & 0.5488  & 0.167  & $\omega$  & 0.7820  &  0.5   \\
\hline
$d\bar d$  & $\eta'$   & 0.9575  & 0.333  &    &    &       \\
\hline
$s\bar s$  & $\eta$   & 0.5488  & 0.667  & $\phi$  & 1.019  & 1   \\
\hline
$s\bar s$  & $\eta'$   & 0.9575  & 0.333  &    &    &       \\
\hline
$c\bar d$  & $D^+$   & 1.869  & 1  & $D^{*+}$  & 2.010  &   1    \\
\hline
$d\bar c$  & $D^-$   & 1.869  & 1  & $D^{*-}$  & 2.010  &   1    \\
\hline
$c\bar u$  & $D^0$   & 1.865  & 1  & $D^{*0}$  & 2.007  &   1    \\
\hline
$u\bar c$  & $\bar{D}^0$   & 1.865  & 1  & $\bar{D}^{*0}$  & 2.007  &   1    \\
\hline
$c\bar s$  & $D_s^+$   & 1.969  & 1  & $D_s^{*+}$  & 2.112  &   1    \\
\hline
$s\bar c$  & $D_s^-$   & 1.969  & 1  & $D_s^{*-}$  & 2.112  &   1    \\
\hline
$c\bar c$  & $\eta_c$   & 2.980  & 1  & $J/\Psi$  & 3.097  &   1    \\
\hline
$u\bar b$  & $B^+$   & 5.279  & 1  & $B^{*+}$  &  5.325  &   1    \\
\hline
$b\bar u$  & $B^-$   & 5.279  & 1  & $B^{*-}$  &  5.325  &   1    \\
\hline
$d\bar b$  & $B^0$   & 5.279  & 1  & $B^{*0}$  &  5.325  &   1    \\
\hline
$b\bar d$  & $\bar{B}^0$   & 5.279  & 1  & $\bar{B}^{*0}$  &  5.325  &   1    \\
\hline
$s\bar b$  & $B_s^0$   & 5.366  & 1  & $B_s^{*0}$  &  5.415  &   1    \\
\hline
$b\bar s$  & $\bar{B}_s^0$   & 5.366  & 1  & $\bar{B}_s^{*0}$  &  5.415  &   1    \\
\hline
$c\bar b$  & $B_c^0$   & 6.594  & 1  & $B_c^{*0}$  &  6.602  &   1    \\
\hline
$b\bar c$  & $\bar{B}_c^0$   & 6.594  & 1  & $\bar{B}_c^{*0}$  & 6.602  &   1    \\
\hline
$b\bar b$  & $\Upsilon$   & 9.460  & 1  &   &   &      \\
\hline
\hline
\end{tabularx}
\label{Meson}
\end{table*}

The mesons and baryons considered are listed in Tables \ref{Meson} and
\ref{Baryon}, respectively. In the tables, the hadron proper probability is
the expectation value (normalization factor) of its quark component wave
function~\cite{lee}. Only the hadron with nonzero proper probability can be
the candidate in the coalescence hadronization. If the coalescing partons have
the same possibility to form a pseudoscalar meson or a vector meson,
(e.g. $u\bar d$ can coalesce into a $\pi^+$ or a $\rho^+$) then the one with
less mass discrepancy between the (invariant) mass of coalescing partons and
the mass of hadron will be preferred. And the same is true for the baryon
production.

\begin{table*}[tbp]
\centering
\renewcommand{\arraystretch}{1.1}
\caption{Baryons in coalescence hadronization model.}
\begin{tabularx}{\textwidth}{C|C|C|C|C|C|C}
\hline
\hline
  & \multicolumn{3}{c|}{Spin-Parity ${\frac{1}{2}}^+$} & \multicolumn{3}{c} {Spin-Parity ${\frac{3}{2}}^+$} \\
 \hline
 Quark conf. & Name & Mass(GeV) &  Proper probability & Name & Mass(GeV) & Proper probability \\
\hline
$ddd$  &   &    &    & $\Delta^-$  & 1.234  &   1    \\
\hline
$ddu$  & $n$   & 0.9396  & 1  & $\Delta^0$  & 1.233  &   1    \\
\hline
$duu$  & $p$   & 0.9383  & 1  & $\Delta^+$  & 1.232  &   1    \\
\hline
$uuu$  &    &  &    & $\Delta^{++}$  & 1.231  &   1    \\
\hline
$dds$  & $\Sigma^-$   & 1.197   &  1  & $\Sigma^{*-}$  & 1.387  &   1    \\
\hline
$dus$  & $\Lambda^0$   & 1.116   &  0.5  &   &  &     \\
\hline
$dus$  & $\Sigma^0$   & 1.193   & 0.5   & $\Sigma^{*0}$  & 1.384  &   1    \\
\hline
$uus$  & $\Sigma^+$   & 1.189   &  1  & $\Sigma^{*+}$  & 1.383  &   1    \\
\hline
$dss$  & $\Xi^-$   & 1.321   &  1  &   $\Xi^{*-}$  & 1.535  &   1    \\
\hline
$uss$  & $\Xi^0$   & 1.315   &  1  &   $\Xi^{*0}$  & 1.532  &   1    \\
\hline
$sss$  &   &   &     &   $ \Omega^- $  & 1.672  &   1    \\
\hline
$ddc$  & $\Sigma_c^0$   & 2.454  & 1  &  $\Sigma_c^{*0}$  & 2.518  &  1   \\
\hline
$duc$  & $\Lambda_c^+$   & 2.284  & 0.5  &   &  &     \\
\hline
$duc$  & $\Sigma_c^+$   & 2.4535  & 0.5  &  $\Sigma_c^{*+}$  & 2.500  &  1   \\
\hline
$dsc$  & $\Xi_c^{0}$   & 2.4703  & 0.5  &    &    &      \\
\hline
$dsc$  & $\Xi_c^{\prime0}$   & 2.550 & 0.5 & $\Xi_c^{*0}$   & 2.630   &  1    \\
\hline
$usc$  & $\Xi_c^{+}$   & 2.4656  & 0.5  &    &    &      \\
\hline
$usc$  & $\Xi_c^{\prime+}$   & 2.550 & 0.5 & $\Xi_c^{*+}$   & 2.630   &   1   \\
\hline
$uuc$  & $\Sigma_c^{++}$   & 2.4529 & 1 & $\Sigma_c^{*++}$   &  2.500  &   1   \\
\hline
$dcc$  & $\Xi_{cc}^+$   & 3.598 & 1 & $\Xi_{cc}^{*+}$    & 3.6565   &   1   \\
\hline
$ucc$  & $\Xi_{cc}^{++}$   & 3.598  & 1 & $\Xi_{cc}^{*++}$    &  3.6565  &   1   \\
\hline
$ssc$  & $\Omega_c^0$   & 2.704  & 1  &  $\Omega_c^{*0}$  &  2.800  &  1    \\
\hline
$scc$  & $\Omega_{cc}^0$   & 3.7866  & 1  &  $\Omega_{cc}^{*0}$  & 3.8247   &   1   \\
\hline
$ccc$  &    &   &   &  $\Omega_{ccc}^{*++}$  & 4.9159   &  1    \\
\hline
$ddb$  & $\Sigma_b^-$    & 5.800  & 1  & $\Sigma_b^{*-}$  & 5.810   &  1    \\
\hline
$uub$  & $\Sigma_b^+$    & 5.800  & 1  & $\Sigma_b^{*+}$  &  5.810  &  1    \\
\hline
$dub$  & $\Sigma_b^0$     & 5.800  & 0.5  & $\Sigma_b^{*0}$ & 5.810   &  1    \\
\hline
$dub$  & $\Lambda_b^0$     & 5.641  & 0.5  &   &    &  1    \\
\hline
$dsb$  & $\Xi_b^{-}$   & 5.840  & 0.5  &    &    &      \\
\hline
$dsb$  & $\Xi_b^{\prime-}$   & 5.960 & 0.5 & $\Xi_b^{*-}$   & 5.970   &   1   \\
\hline
$usb$  & $\Xi_b^{0}$   & 5.840  & 0.5  &    &    &      \\
\hline
$usb$  & $\Xi_b^{\prime0}$   & 5.960 & 0.5 & $\Xi_b^{*0}$   &  5.970  &   1   \\
\hline
$dcb$  & $\Xi_{bc}^{0}$   & 7.0057  & 0.5  &    &    &      \\
\hline
$dcb$  & $\Xi_{bc}^{\prime0}$   & 7.0372 & 0.5 & $\Xi_{bc}^{*0}$   & 7.0485   &   1   \\
\hline
$ucb$  & $\Xi_{bc}^{+}$   & 7.0057  & 0.5  &    &    &      \\
\hline
$ucb$  & $\Xi_{bc}^{\prime+}$   & 7.0372 & 0.5 & $\Xi_{bc}^{*+}$   & 7.0485   &   1   \\
\hline
$dbb$  & $\Xi_{bb}^{-}$   & 10.4227  & 1  &  $\Xi_{bb}^{*-}$  & 10.4414   &      \\
\hline
$ubb$  & $\Xi_{bb}^{0}$   & 10.4227  & 1  &  $\Xi_{bb}^{*0}$  & 10.4414   &      \\
\hline
$ssb$  & $\Omega_b^-$   & 6.120  & 1  &  $\Omega_b^{*-}$  & 6.130   &  1    \\
\hline
$scb$  & $\Omega_{bc}^0$   & 7.191  & 0.5  &    &    &   1   \\
\hline
$scb$  & $\Omega_{bc}^{\prime0}$   & 7.211  & 0.5  &  $\Omega_{bc}^{\prime*0}$  & 7.219   &   1   \\
\hline
$sbb$  & $\Omega_{bb}^-$   & 10.6021  & 1  &  $\Omega_{bb}^{*-}$  & 10.6143   &   1   \\
\hline
$ccb$  & $\Omega_{bcc}^{+}$   & 8.3095  & 1  &  $\Omega_{bcc}^{*+}$  &  8.3133  &   1   \\
\hline
$cbb$  & $\Omega_{bbc}^{0}$   & 11.7077  & 1  &  $\Omega_{bbc}^{*0}$  &  11.7115  &   1   \\
\hline
$bbb$  &    &   &    &  $\Omega_{bbb}^{*-}$  & 15.1106   &   1   \\
\hline
\hline
\end{tabularx}
\label{Baryon}
\end{table*}

\section{Comparison with experiments} \label{sec:comparison}
\subsection{High energy reaction}
In B- and C-loop simulations, if the hadronization is implemented by LSF, the
key parameters are $K$, $\sigma_G$, $\alpha$, and $\beta$ (`adj1(10)',
`adj1(34)', `adj1(6)' and `adj1(7)' in PACIAE, corresponding to PARP(31),
PARJ(21), PARJ(41) and PARJ(42) in PYTHIA). $K$ is a multiplicative factor of
hard scattering cross sections as shown in Eq.~(\ref{qqcro}). $\sigma_G$ is
the width of Gaussian $p_x$ and $p_y$ transverse momentum distributions for
the primary hadrons~\cite{PYTHIA}. $\alpha$ and $\beta$ are the parameters in
the LUND fragmentation function~\cite{PYTHIA,Andersson:1983jt}:
\begin{equation}
    f(z) \propto z^{-1} (1-z)^\alpha \exp(-\beta m_T^2/z),
\end{equation}
where $z$ is the fraction of energy taken by a hadron fragmented from a parton
and $m_T^2 = m^2 + p_T^2$ is transverse mass of the hadron. The $\sigma_G$,
$\alpha$ and $\beta$ hence couple with each other.

On the other hand, if the coalescence hadronization model (Coal) is selected
in C-loop (note: B-loop is hadronized by LSF only), the key parameters would be
$K$, $\sigma_q$, $e_{she}$ (`adj1(10)', `adj1(34)', and adj1(7)), and
`adj1(16)'. Here $K$ has the same meaning as mentioned above. $\sigma_q$
and $e_{she}$ are the width of excited quark-antiquark $p_T$ in the energetic
quark deexcitation and the threshold energy of deexcitation, respectively.
The `adj1(16)' refers to the allowed maximum number of deexcitation generation
(D=1).

The midrapidity charged particle multiplicity density $dN_{ch}/d\eta$ are given
in Table.~\ref{tab:eta_density}. Here we see the PACIAE model results well
reproduce the experimental data from PHOBOS~\cite{PHOBOS:2010eyu} and
ALICE~\cite{ALICE:2010mlf}.

\begin{table*}
    \caption{The midrapidity charged particle
             multiplicity density $dN_{ch}/d\eta$ in 0-6\% most central Au+Au
             collisions at $\sqrt{s_{NN}} = 0.2$~TeV for
             $|\eta| < 1$ from PHOBOS~\cite{PHOBOS:2010eyu} and 0-5\% most
             central Pb+Pb collisions at $\sqrt{s_{NN}} = 2.76$~TeV for
             $|\eta| < 0.5$ from ALICE~\cite{ALICE:2010mlf} compared with the
             results from PACIAE 3.0 B-loop, C-loop LSF and C-loop Coal.}
    \label{tab:eta_density}
    \begin{ruledtabular}
    \begin{tabular}{cccccc}
    System & $\sqrt{s_{NN}}$ (TeV) & Exp. & B-Loop & C-loop LSF & C-loop Coal \\
    \hline
    Au+Au & 0.2
          & $1310 \pm 69$~\footnote{ Taken from PHOBOS~\cite{PHOBOS:2010eyu} }
          & 1283 & 1301 & 1288 \\
    Pb+Pb & 2.76
          & $1610 \pm 60$~\footnote{ Taken from ALICE~\cite{ALICE:2010mlf} }
          & 1667 & 1672 & 1587
    \end{tabular}
    \end{ruledtabular}
\end{table*}

\begin{figure}[htbp]
\centering
\includegraphics[width=0.44\textwidth]{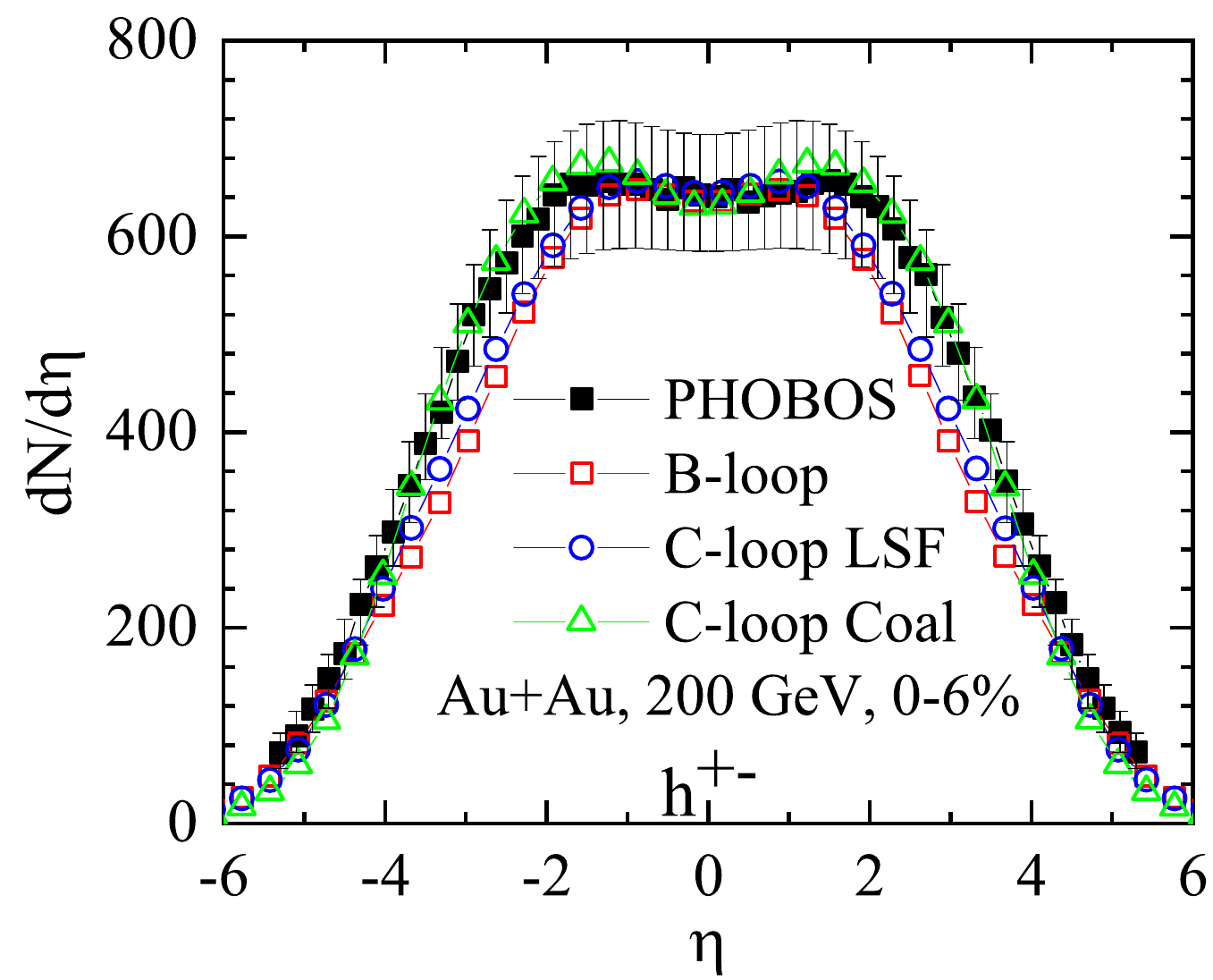}
\caption{(Color online) Charged particle pseudorapidity distribution in
    0-6\% most central Au+Au collisions at $\sqrt{s_{NN}}=$ 200 GeV from
    PACIAE model simulations comparing with PHOBOS data~\cite{Back:2002wb}.}
\label{fig:Au200_dNdeta}
\end{figure}

\begin{figure}[htbp]
\centering
\includegraphics[width=0.44\textwidth]{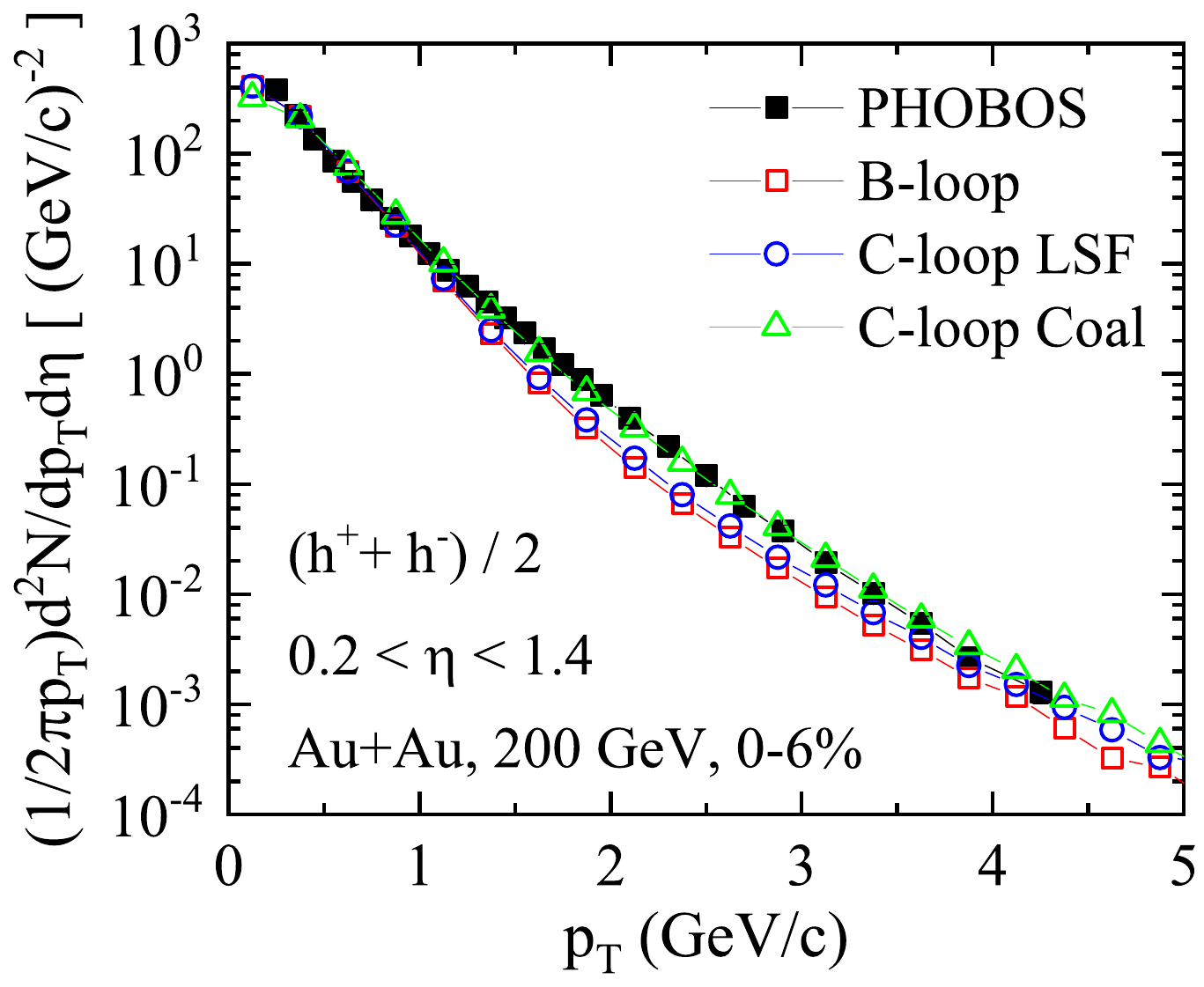}
\caption{(Color online) Charged particle invariant transverse momentum
    spectra in 0-6\% most central Au+Au collisions at $\sqrt{s_{NN}}=$ 200 GeV
    from PACIAE simulations comparing with PHOBOS data~\cite{PHOBOS:2003wxa}.}
\label{fig:Au200_inv_pT}
\end{figure}

In Fig.~\ref{fig:Au200_dNdeta}, we compare the PHOBOS charged particle
pseudorapidity distribution~\cite{Back:2002wb,PHOBOS:2003wxa} (black solid
squares) measured in 0-6\% most central Au+Au collisions at
$\sqrt{s_{NN}}=$200 GeV with B- and C-loop simulation results. The results of
B-loop are indicated by red open squares, while the C-loop LSF and Coal are,
respectively, indicated by blue open circles and green open triangles.
Fig.~\ref{fig:Au200_inv_pT} is the same as Fig.~\ref{fig:Au200_dNdeta} but
for the transverse-momentum spectrum. In the simulations, the parameters were
tuned as follows:
\begin{itemize}
    \item[(1)] B-loop: $K=0.9, \sigma_G=0.45, \alpha=0.3, \beta=0.58$.
    \item[(2)] C-loop LSF: $K=2.5, \sigma_G=0.45, \alpha=0.3, \beta=0.1,$ and
                $\rm{PARP(82)} = 2.5$~\footnote{ The regularization scale of
                transverse-momentum spectrum for multiple interactions,
                parameter parp82 in PACIAE. }.
    \item[(3)] C-loop Coal: $K=0.7, \sigma_q=0.6, e_{she}=1.8$, and
                $\rm{PARP(91)} = 1.3$~\footnote{ The width of primordial
                transverse momentum $k_\perp$ for the partons inside the beam
                hadrons, parameter adj1(39) in PACIAE. }.
\end{itemize}
One can see in these two figures that the PACIAE model well reproduces the
PHOBOS data within the error bars.

\begin{figure}[htbp]
\centering
\includegraphics[width=0.44\textwidth]{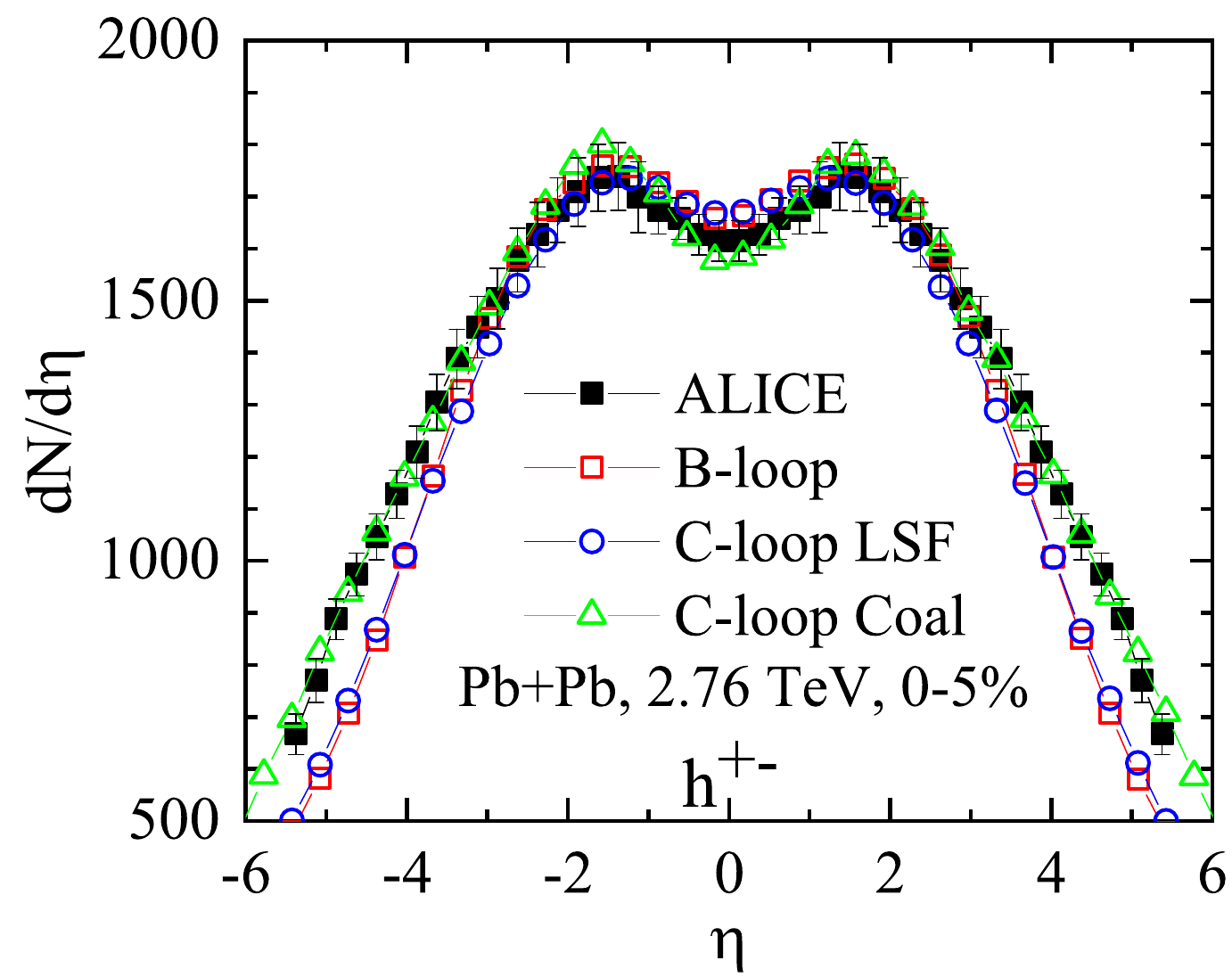}
\caption{(Color online) Charged particle pseudorapidity distribution in
    0-5\% most central Pb+Pb collisions at $\sqrt{s_{NN}}=$ 2.76 TeV from
    PACIAE simulations comparing with ALICE data~\cite{ALICE:2013jfw}.}
\label{fig:Pb2760_dNdeta}
\end{figure}

\begin{figure}[htbp]
\centering
\includegraphics[width=0.44\textwidth]{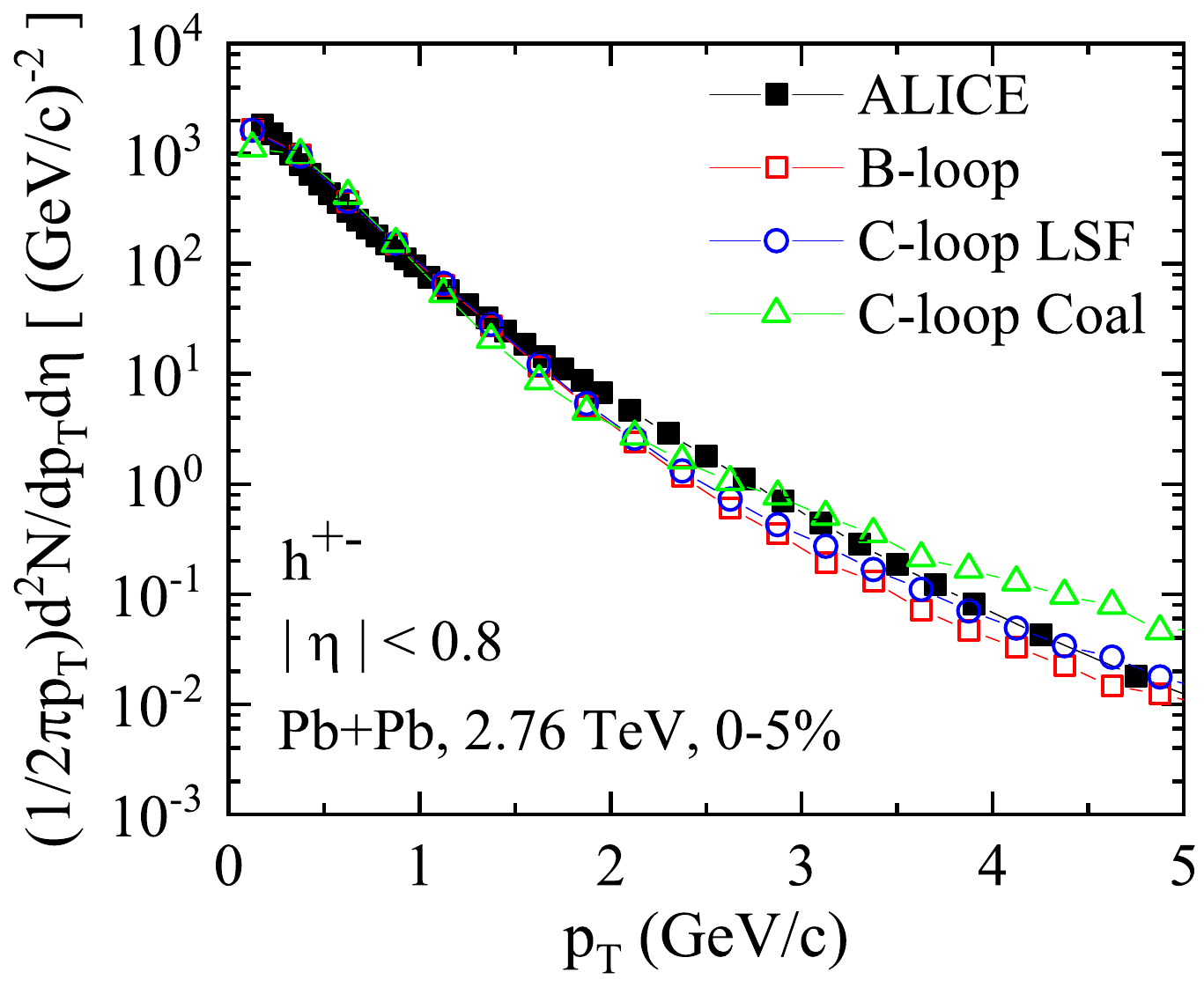}
\caption{(Color online) Charged particle invariant transverse momentum
    spectra in 0-5\% most central Pb+Pb collisions at $\sqrt{s_{NN}}=$ 2.76 TeV
    from PACIAE simulations comparing with ALICE data~\cite{ALICE:2012aqc}.}
\label{fig:Pb2760_inv_pT}
\end{figure}

A similar comparison with ALICE data measured in 0-5\% most central Pb+Pb
collisions at $\sqrt{s_{NN}}=$ 2.76 TeV is shown in
Figs.~\ref{fig:Pb2760_dNdeta} and \ref{fig:Pb2760_inv_pT}. The parameters are:
\begin{itemize}
    \item[(1)] B-loop: $K=2.9, \sigma_G=0.6, \alpha=0.3, \beta=0.13$.
    \item[(2)] C-loop LSF: $K=2.9, \sigma_G=0.6, \alpha=0.3, \beta=0.012$.
    \item[(3)] C-loop Coal: $K=1.5, \sigma_q=0.6, e_{she}=1.9$, and
               $ \rm{PARP(91)} = 0.6$.
\end{itemize}
Figs.~\ref{fig:Pb2760_dNdeta} and ~\ref{fig:Pb2760_inv_pT} show that, PACIAE
model gives good descriptions to the ALICE charged particle pseudorapidity
distribution~\cite{ALICE:2013jfw} and $p_T$ distribution
data~\cite{ALICE:2012aqc}, except that the $p_T$ distribution from 
``C-loop Coal'' is slightly harder at $p_T > 4$ GeV/c region.

\subsection{Low energy reaction}
In the PACIAE A-loop simulation there are two parameters only. One is the
ratio of the inelastic to total cross section $R_{inela/tot}$
(`x\_ratio' in program), another is the instantaneous decay probability of
$\Delta$ particle (`decpro' in program). The $R_{inela/tot}$ is assumed to be
a function of the incident channel $\sqrt{s_{NN}}$~\cite{Bertsch,Cugnon}:
\begin{equation}
R_{inela/tot}=\frac{1.35(\sqrt{s_{NN}}-2.015)^2}
               {0.015+(\sqrt{s_{NN}}-2.015)}, \\
               \rm{if}~ \sqrt{s_{NN}} < 3~\rm{GeV}.
\end{equation}

\begin{figure}[htbp]
\centering
\includegraphics[width=0.55\textwidth]{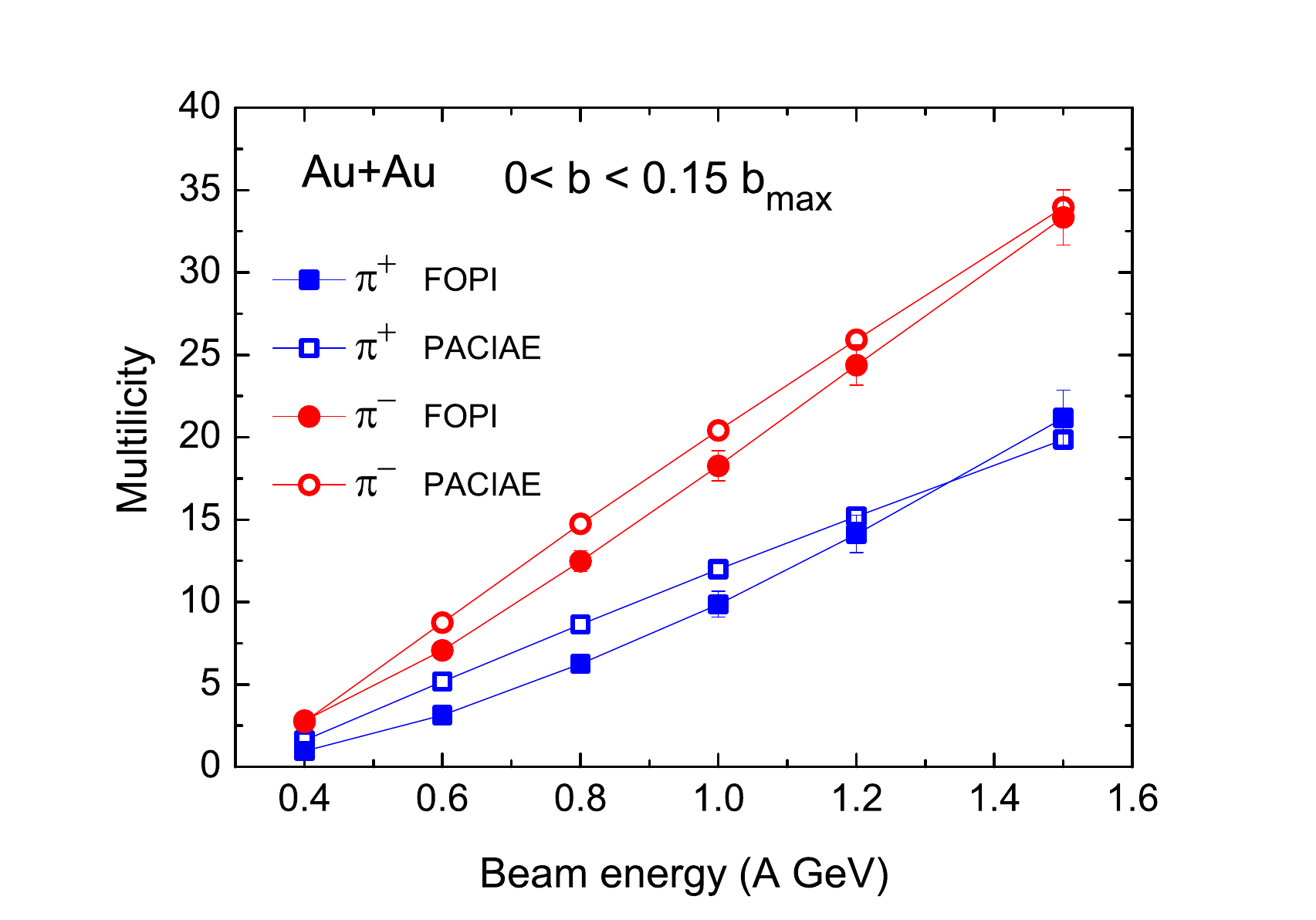}
\caption{(Color online) The $\pi^+$ and $\pi^-$ yields in most central Au+Au
collisions at beam energy of 0.40, 0.60, 0.80, 1.0, 1.2, and 1.5A GeV from
PACIAE model simulations comparing with the corresponding FOPI experimental
data~\cite{Reisdorf}.}
\label{multi}
\end{figure}

In Fig.~\ref{multi}, we compare PACIAE simulated results (decpro=0.9) of
$\pi^+$ and $\pi^-$ yields to the corresponding FOPI experimental
data~\cite{Reisdorf} in most central Au+Au collisions at beam energy (fixed
target) of 0.40, 0.60, 0.80, 1.0, 1.2, and 1.5A GeV (corresponding to
$\sqrt{s_{NN}}$ equal to 2.066, 2.155, 2.241, 2.402, and 2.520A GeV,
respectively). Here one sees the results of PACIAE model well reproduce the
experimental data.

\begin{figure}[htbp]
\centering
\includegraphics[width=0.55\textwidth]{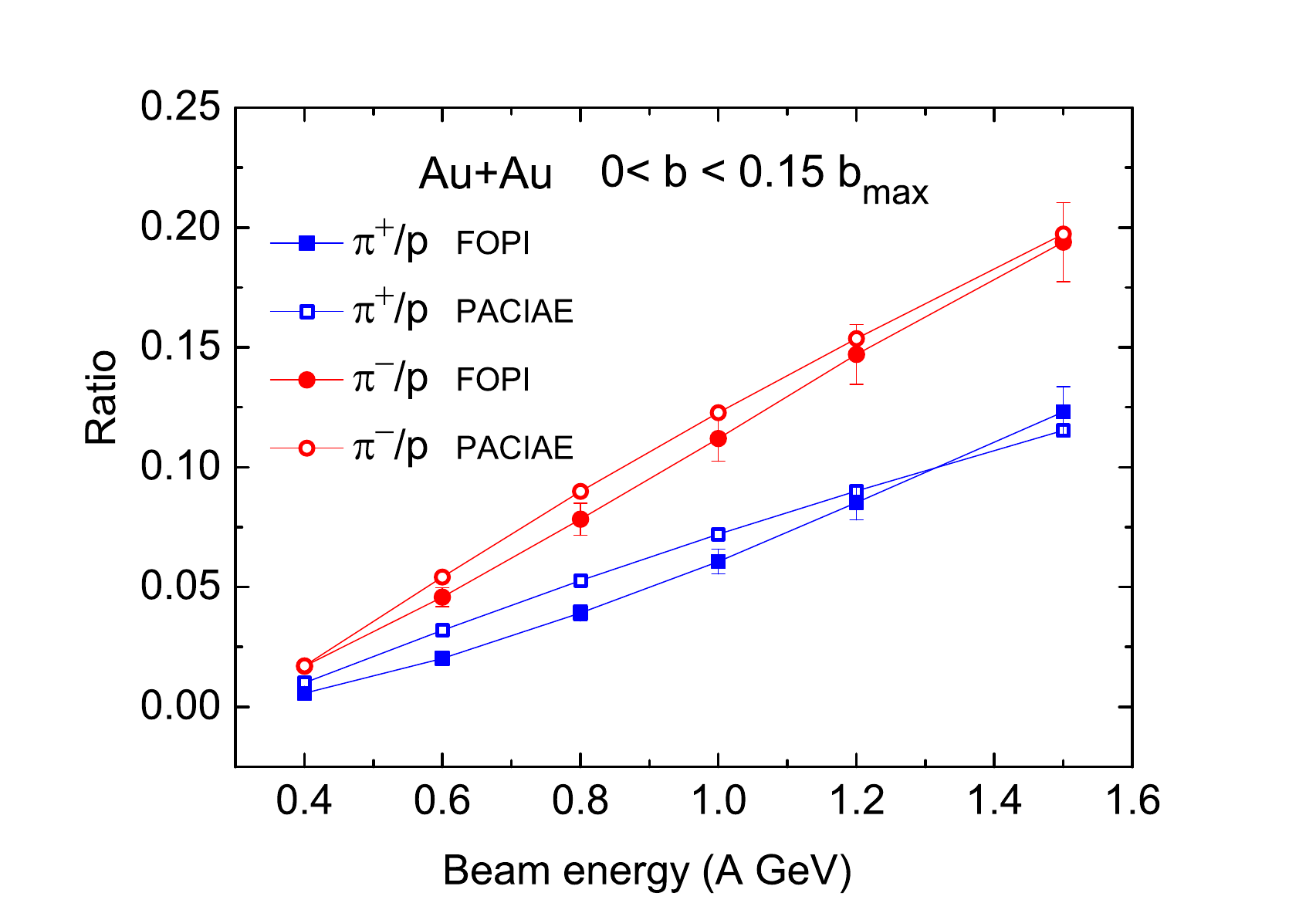}
\caption{(Color online) The $\pi^+/\rm{p}$ and $\pi^-/\rm{p}$ ratios in
most central Au+Au collisions at beam energy of 0.40, 0.60, 0.80, 1.0, 1.2,
and 1.5A GeV from PACIAE model simulations comparing with the FOPI
experimental data~\cite{Reisdorf}.}
\label{ratio}
\end{figure}

The PACIAE model results of $\pi^+/\rm{p}$ and $\pi^-/\rm{p}$ ratios are shown
in Fig.~\ref{ratio} and compared with FOPI experimental data measured in the
same collision system like Fig.~\ref{multi}. Since in the final
hadronic state generated in the PYTHIA (PACIAE) model the light nuclei (d, t,
$^3\rm{He}$, $^4\rm{He}$, Li, etc.) are not identified. The charge number of
above light nuclei must first be added into the proton data and then compared
with PACIAE results due to the charge conservation principle. Fig.~\ref{ratio}
shows the PACIAE results reproduce the FOPI experimental data generally well.

\begin{figure}[htbp]
\centering
\includegraphics[width=0.55\textwidth]{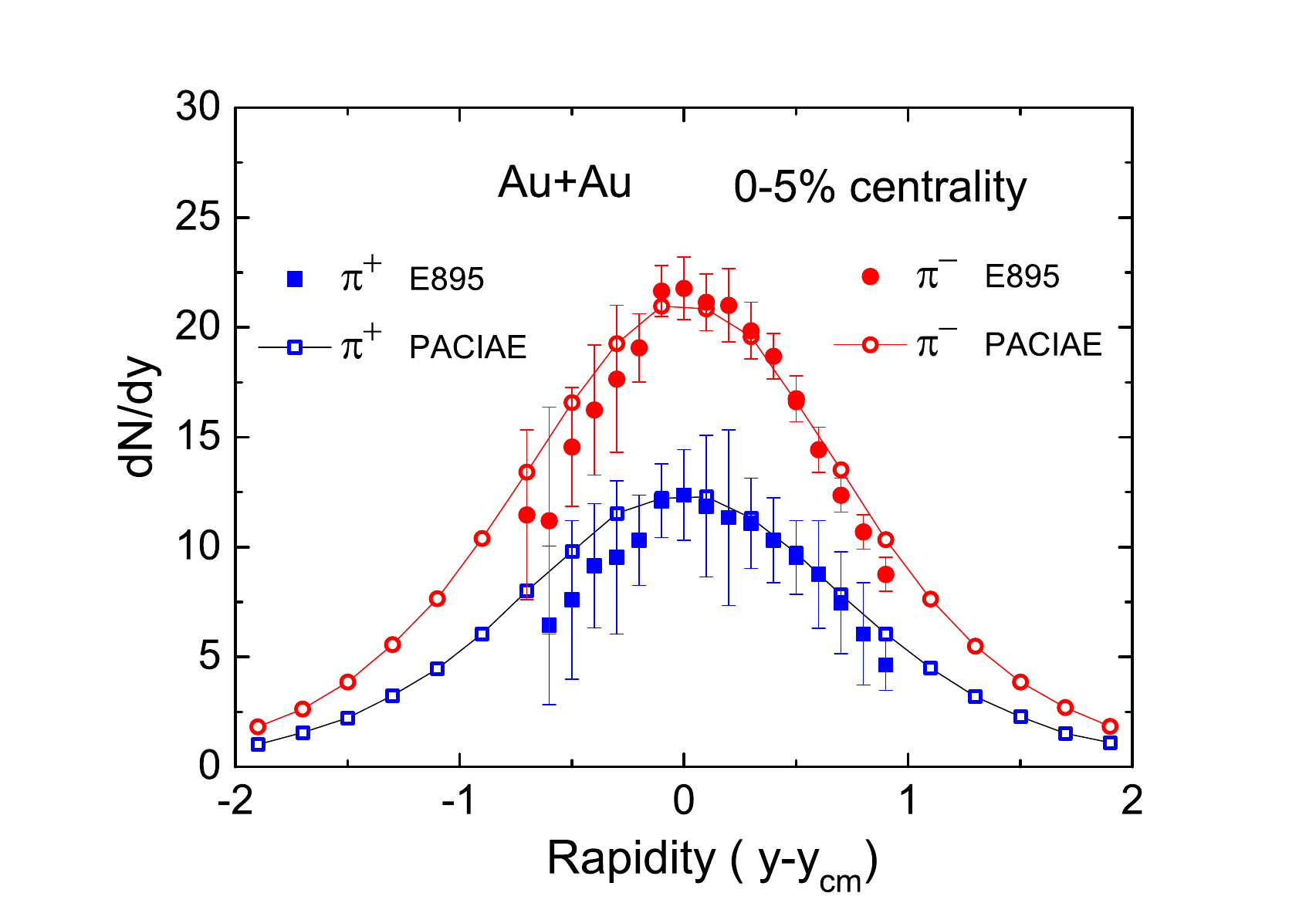}
\caption{(Color online) The E895 $\pi^+$ and $\pi^-$ experimental rapidity
distributions in 0-5\% most central Au+Au collisions at 1.85A GeV actual beam
energy~\cite{Klay} comparing with PACIAE model simulations.}
\label{eta}
\end{figure}

The E895 measured $\pi^+$ and $\pi^-$ rapidity distributions~\cite{Klay} in
0-5\% most central Au+Au collisions at nominal beam energy of 2 GeV/nucleon
are compared with PACIAE results in Fig. \ref{eta}. The actual beam energy
after correction for the energy loss is 1.85A GeV, which is corresponding to
$\sqrt{s_{NN}}$ = 2.64 GeV. One can see here the E895 measured $\pi^+$ and
$\pi^-$ rapidity distributions are fairly well reproduced by PACIAE.

\begin{figure}[htbp]
\centering
\includegraphics[width=0.55\textwidth]{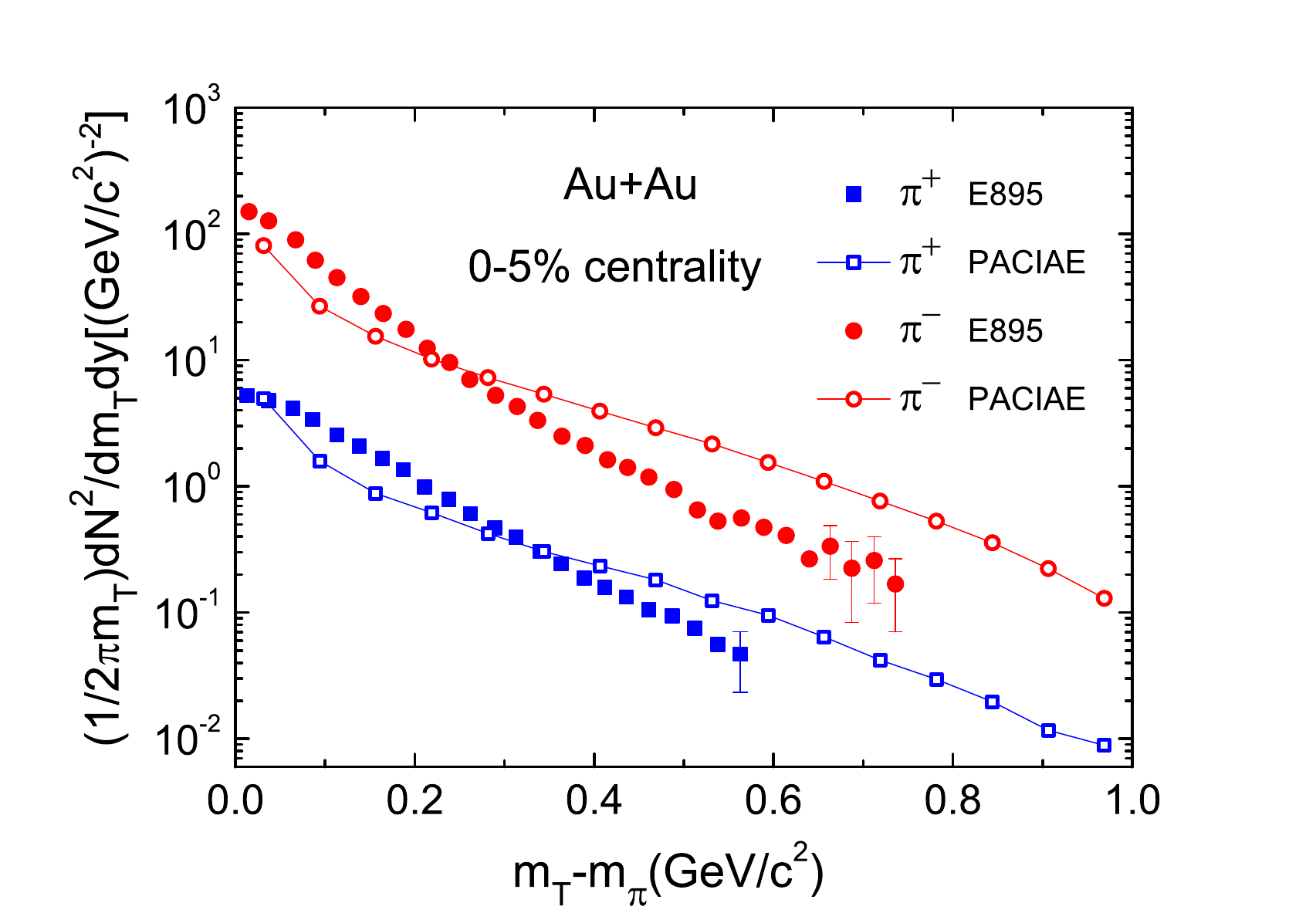}
\caption{(Color online) The same as the Fig.~\ref{eta} but for $\pi^+$ and
$\pi^-$ transverse mass distributions.}
\label{pt}
\end{figure}

Similarly, Fig. \ref{pt} gives the comparison of E895 measured $\pi^+$ and
$\pi^-$ transverse mass (transverse momentum) distributions~\cite{Klay} to the
PACIAE results in the same collision system like Fig. \ref{eta}. One can
see here the experimentally measured $\pi^+$ and $\pi^-$ transverse
mass distributions are harder than PACIAE simulations in the transverse mass
interval of 0.1-0.3 GeV$/c^2$, otherwise softer than PACIAE results. As the
outgoing particle momentum in low energy A-loop simulation is fixed by the
two-body scattering kinematic, and there are no adjustable parameters unlike
that in the high energy B- and C-loop simulations, the improvement of the
agreement between experiment and theory in particle transverse mass
distribution has to be studied further.

\section{Conclusion}
We have constructed a phenomenological parton and hadron cascade model
PACIAE 3.0 based on PYTHIA 6.428 and PACIAE 2.2 series. The C-, B- and A-loop
simulations are designed for the high and low energy nuclear collisions,
respectively. In C-loop simulation, the parton-parton inelastic scatterings
are implemented. The single string structure and multiple string interaction
mechanisms are involved in the high energy B- and C-loop simulations for the
strangeness enhancement investigation. An improved mapping formula between the
percentage and impact parameter centrality definitions is proposed responding
the ALICE, ATLAS, and CMS observation of $b_{max}\approx 20$ fm. The
phenomenological coalescence hadronization model is also extensively modified
in the C-loop simulation.

The simulated results are compared with the experimental data measured in FOPI
and E895 experiments, and at RHIC as well as LHC energies, respectively.
Generally speaking, the basic experimental data of particle yield, transverse
momentum distribution, and the pseudorapidity distribution are reasonably
reproduced.

It seems necessary to introduce the mean field, Fermi motion, and Pauli
blocking effects in the A-loop simulation for the study of symmetry energy and
the equation of state. For the investigation of heavy flavor production in
relativistic nuclear collisions with B- and/or C-loop simulations, it may be
obliged to open the special channels for `Heavy flavours' sector in PYTHIA. It
is a bias sampling method, the calculated results must be multiplied by a
correcting (normalization) factor before comparison with experimental data.

At last, from a technical point of view, PACIAE 3.0 is written in FORTRAN
programming language and based on PYTHIA 6. With
the development of physics and computer science, high-energy community
embraces more modern languages and technologies, in particular from FORTRAN to
object-oriented C++ language. A plan of accessing to C++-based
PYTHIA 8~\cite{pythia83} is one of our future goals and is on the timetable, in 
which we expect more fruitful physics to be integrated with PACIAE.

\section*{Acknowledgements}
The author An-Ke Lei thanks Xiao-Ming Zhang, Shu-Su Shi, and Wen-Chao Zhang
for helpful discussions.
This work was supported by the National Natural Science
Foundation of China (Grandts Nos.: 12375135, 11775094, 11905188, 12275322) and 
the 111 project of the foreign expert bureau of China. Yu-Liang Yan 
acknowledges the financial support from Key Laboratory of Quark and Lepton 
Physics in Central China Normal University (Grant No. QLPL201805) and the 
Continuous Basic Scientific Research Project (Grant No. WDJC-2019-13).

\appendix
\section{PACIAE 3.0 user's guide}

\subsection{Program running}
To run PACIAE 3.0, one direct way is to compile the source code, modify the
input file ``usu.dat'' as needed, and execute the program. Another way is to
use the toy SHELL-script ``PACIAE.sh''. A Makefile has been integrated in the
``PACIAE.sh'' with GFortran compiler specified. It will compile the
source code, generate ``usu.dat'' (the old ``usu.dat'' will be overwritten)
and run the program automatically. More details could be found in
``README.md'' file.

PACIAE 3.0 comes with a simple internal on-line analyzing module and outputs
several files. The analyzing output file is ``rms.out'', where some basic
results of collisions and six distributions (rapidity distribution $dN/dy$,
invariant transverse momentum spectrum $1/p_T dN/dp_T$, pseudorapidity
distribution $dN/d\eta$, invariant transverse mass spectrum $1/m_T dN/dm_T$,
event-wise multiplicity distribution, and transverse momentum
spectrum $dN/dp_T$) are provided. The ``rms0.out'' is a file recording the
input parameters. The ``main.out'' file is PYTHIA-style particle list output
file. If user chooses to output OSCAR-format file, there will be an
``oscar.out'' file that records list of final state particles or full
event history.

\subsection{The basic tuning criteria}
In Sec.~\ref{sec:comparison}, we have given rough tuning results at both low-
and high-energies. A ``tune'' essentially requires a very large amount of
experimental data fitting with a couple of adjustment parameters, such as the
Perugia 2011 tune of PYTHIA 6~\cite{Skands:2010ak} and Monash 2013 tune of
PYTHIA 8~\cite{,Skands:2014pea} the ALICE typically used. However, for
heavy-ion collisions, it is impossible to meet a ``perfect'' tune due to our
inadequate understanding of this very sophisticated large system. A recommended
effective tuning criterion is: Fit the midrapidity density, pseudorapidity
distributions and/or transverse momentum spectra of basic charged particles to
the experimental data at the corresponding system and energy. Then one could
conduct other studies of interest. Another criterion is based on what one
would like to study. For instance, to study the topic of nuclear modification
factors $R_{AA}$, one can fit the $R_{AA}$ of $\pi^\pm$ to experimental data
at first~\cite{Sa:2022pnd}.

\subsection{Incident channel selection in the update of $hh$ collision list}
The particle yield in final hadronic state is sensitively depended on the
selection of incident channel in the update of $hh$ collision list after each
$hh$ collision. Presently, only the $NN$ collision is selected at the beginning
of $hh$ simulation loop and in the subroutine of `updtlp', `updatl', and
`intdis' consistently, in the B- and C-loop simulations. In A-loop simulation,
only the $NN$, $\Delta N$, and $\pi N$ are selected at the beginning of $hh$
collision simulation loop and in the subroutine of `updatl\_{nn}' and `intdis'
consistently.

\subsection{Main switches and parameters}
In the follows we list main switches and parameters as well as their
potentials, respectively, for user reference. As mentioned above the `decpro'
and `x\_ratio' are the only two free parameters in the A-loop simulation, thus
the following Table~\ref{switch} is just for the B- and C-loop
simulations only. More details could be found in ``usu.dat'', ``PACIAE.sh'',
and the comments in ``main\_30.f''.

\begin{table*}[tbp]
\caption{Object of study vs. switch and/or parameter.}
\setlength{\tabcolsep}{14.5pt}
\renewcommand{\arraystretch}{1.8}
\begin{tabular}{c|c}
\hline
\hline
Simulation mode & iMode~\footnote{name in program (the same later)}
                  = 1: A-loop ; =2: B-loop ; =3: C-loop. \\
\hline
QCD sub-processes selection &
\makecell{ nchan = 0: inelastic (INEL); = 1: non-single diffractive (NSD);
                 = 2: Drell-Yan; \\
                 = 3: $J/\Psi$ production; = 4: heavy-flavor production;
                 = 5: direct photon; = 6: soft only; \\
                 = 7: $W^{+/-}$ production; = 8: default PYTHIA;
                 = 9: $Z^{0}$ production. } \\
\hline
Chiral magnetic effect & adj1(3)=0: off; =1, on. \\
\hline
Hadronization model & adj1(12) = 0: string fragmentation model;
                               = 1: coalescence model. \\
\hline
Parton rescattering &
    \makecell{ adj1(1) $>$ 0: with; = 0: without. \\
               adj1(1) is a factor multiplying on
               parton-parton cross section. } \\
\hline
Process in parton rescattering &
    \makecell{ iparres = 0: elastic processes only; \\
                       = 1: elastic + inelastic processes. \\
           i\_inel\_proc = 6: with inelastic process of 4, 6, and 7; \\
                       = 7: with inelastic process 7 only. } \\
\hline
String fragmentation model &
 \makecell{ adj1(10): K factor, adj1(6): $\alpha$, adj1(7): $\beta$, adj1(34):
$\sigma_H$. } \\
\hline
Coalescence model &
 \makecell{ adj1(16): allowed number of deexcitation generation,\\
            adj1(17): threshold energy of deexcitation, \\
            adj1(29): deexcitation function, \\
            adj1(34): $\sigma_q$. } \\
\hline
Effective string tension &
 \makecell{ kjp22 = 1: variable single string tension; \\
                  = 2: variable multiple string tension; \\
                  = 3: variable single + multiple string tension; \\
                  = 4: constant string tension. } \\
\hline
Hadron rescattering & kjp21 = 1: with;  =0: without. \\
\hline
\hline
\end{tabular}\label{switch}
\end{table*}

\end{document}